\documentclass[prd,aps,showpacs,tightenlines,10pt,reprint,nofootinbib,superscriptaddress, floatfix,twocolumn]{revtex4-2}

\usepackage[utf8]{inputenc}
\usepackage{amssymb}
\usepackage{tabularx}
\usepackage{multirow}
\usepackage{hyperref}
\usepackage{amsmath}
\usepackage{amsthm}  
\usepackage[left=2.5cm,right=2.5cm,top=2.5cm,bottom=2.5cm]{geometry}
\usepackage{tensor}
\usepackage{mathrsfs}
\usepackage{enumitem}
\usepackage{float} 
\usepackage{lipsum}
\usepackage{comment}
\usepackage[dvipsnames]{xcolor} 
\usepackage{graphicx}
\usepackage{orcidlink} 
\usepackage[normalem]{ulem} 
\newcommand{\CJ}[1]{\textcolor{black}{#1}}

\newcommand{\LM}[1]{\textcolor{black}{#1}}
\newcommand{\LMtwo}[1]{\textcolor{black}{#1}}

\newcommand{\nablastar}{\overset{\star}{\nabla}}

\newcommand{\LL}{\ensuremath{\mathcal{L}}}

\newcommand{\ddelta}[2]{\frac{\delta #1}{\delta #2}}

\renewcommand{\eqref}[1]{Eq.~(\ref{#1})}

\newcommand{\e}[1]{\cdot 10^{#1}} 

\let\oldphi\varphi \let\varphi\phi \let\phi\oldphi
\let\oldepsilon\varepsilon \let\varepsilon\epsilon \let\epsilon\oldepsilon

\begin{document}

\title{Effect of Torsion on Neutron Star Structure in Einstein-Cartan Gravity}

\author{Cédric Jockel\,\orcidlink{0009-0007-7617-7178}}
\email{cedric.jockel@aei.mpg.de}
\affiliation{Max Planck Institute for Gravitational Physics (Albert Einstein Institute), Am Mühlenberg 1, 14476 Potsdam, Germany, European Union}
\affiliation{Institute for Theoretical Physics, Goethe University, 60438 Frankfurt am Main, Germany}

\author{Leon Menger\,\orcidlink{0000-0003-0621-4977}}
\email{leonmenger@nd.edu}
\affiliation{Department of Mathematics, Notre Dame University, 255 Hurley Bldg, Notre Dame, IN 46556, USA}
\affiliation{Department of Physics, ETH Zürich, Rämistrasse 101, 8092,
Zürich, Switzerland}

\date{\today}

\begin{abstract}
Einstein--Cartan gravity is a close historical sibling of general relativity that allows for spacetime torsion. As a result, angular momentum couples to spacetime geometry in a similar way to energy.
While consequences of this are well studied on cosmological scales, their role in neutron star physics is largely under-explored.
We study the effects that torsion, sourced by either microphysical spin or macroscopic angular momentum, has on neutron stars. For this, we use a simplified polytropic model to quantify the microphysical coupling to torsion. We also derive expressions to model rotation-induced torsion effects and estimate the consequences for rotating neutron stars with different rotation rates.
We find that the presence of torsion in general leads to neutron stars with smaller radii and masses, but higher central densities. Realistic models for microphysical spin lead to torsion effects that have no relevant influence on the neutron star structure. Rotation-induced torsion effects however, can decrease the radius by up to $900\,m$, which is comparable to the increase due to centrifugal forces. Depending on which effect dominates, this leads to a torsion-induced spin-up or spin-down of the neutron star.
We conclude that torsion effects due to rotation can not be neglected and are large enough to be tested using current or near-future technology.
\end{abstract}

\maketitle

\section{Introduction \label{sec:intro}} 
General relativity (GR) stands as one of the most successful physical theories to date and has withstood a multitude of experimental tests \LM{(see \cite{Will:2014kxa,Berti:2015itd})}. As a result, it is widely accepted as the predominant theory of gravity. It does however leave several open questions about the nature of cosmological evolution and dark energy as well as the failure to derive a quantized theory of gravity. GR also has unreconciled singularities in the centre of black holes and at the big bang. A vast amount of \LMtwo{alternatives and equivalents} to general relativity have been put forward to patch these shortcomings (see \cite{Clifton:2011jh} for a review). Historically one of the first is Einstein--Cartan (EC) gravity\footnote{For an interesting historic recount of letters between Cartan and Einstein, see \cite{Cartan_Einstein_Letters}}, which is the subject of this work (for a review see \cite{HehlReview76}). \LMtwo{It is completely equivalent to GR in vacuum. However when matter is present, the two theories differ.} \\

In the leading paradigm to describe gravity, curvature is used to describe how matter and spacetime interact. However, curvature is only one of three possible geometric properties of spacetime, which are determined by the spacetime connection. Given a metric $g$, a connection $\nabla$ has three purely geometric properties: Curvature, torsion and non-metricity. Curvature describes the failure of the associated parallel transport to commute, i.e. $\nabla_\mu\nabla_\nu \neq \nabla_\nu\nabla_\mu$. Torsion describes the failure of parallel transport parallelograms to close. Non-metricity corresponds to the failure of the covariant derivative of the metric to vanish, i.e. $\nabla g \neq 0$. It is related to the (invariance of the) measured length under parallel transport. \\
The Levi--Civita connection $\nabla_{LC}$ is the unique connection such that there is no torsion and $\nabla g = 0$. The connection coefficients of $\nabla_{LC}$, the Christoffel symbols, are symmetric in their lower two indices, i.e. $\Gamma^{\sigma}_{\mu\nu} = \Gamma^{\sigma}_{\nu\mu}$. The torsion of a connection can be expressed as the antisymmetric part of the connection coefficients. Even if we assume metric-compatibility -- which we do throughout this work -- a general connection can have both curvature and torsion. \\

In general relativity, the Levi--Civita connection is assumed a priori. This leads to spacetime being described only by its curvature. There is however no inherent physical reason, why curvature is chosen to describe spacetime. In fact, there exist entirely equivalent formulations of general relativity which only use torsion \cite[section III]{ARCOS_2004} \cite[section 4]{Hammond2002} (this is commonly called the teleparallel equivalent to general relativity) or only non-metricity \cite{heisenberg2023review} as dynamical fields (also see \cite{BeltranJimenez:2019esp} for a review on this topic). A more general way of formulating these choices is metric-affine gravity. There, the connection is chosen as an independent degree of freedom alongside the metric. EC gravity can also be considered as a special case of metric affine gravity. \\
The preference of using only curvature to describe gravity has thus mainly historical reasons. It is important to stress that the a priori exclusion of torsion has no inherent physical or observational reason. \\

In this paper we consider the case where gravity is described by both curvature and torsion \LM{and where matter couples minimally to gravity}. The simplest such theory is Einstein--Cartan (EC) gravity. It was historically derived as a way to include \LM{the concept of torsion into the theory of gravity by Cartan \cite{Cartan1923, Cartan1924a, Cartan1924b}. Kibble and Sciama \cite{Kibble:1961ba, Sciama:1964wt} found an alternative derivation of EC theory by gauging the Poincaré group. It was later shown that EC is the most general gauge theory of the Poincaré group (up to topological terms), see e.g. \cite{Gronwald:1995em}.} Here the translational part of the gauge group leads to a conservation of energy-momentum. It couples to spacetime via curvature. Conversely, the rotational part of the group corresponds to a conservation o+f spin angular momentum. It couples to spacetime via torsion (also see \cite{HehlReview76} for a review). General relativity also emerges as the zero-torsion limit of this theory, see \cite{Obukhov:2020uan}. \\

Einstein--Cartan gravity in total has two geometric degrees of freedom: Curvature and torsion. It therefore also features two sets of field equations, which describe the dynamics of these spacetime properties. \\
One of the most prominent \LM{potential} consequences of allowing torsion is \LM{the prospect of} singularity-avoidance. It was \LM{proposed} that, in big-bang and black hole formation scenarios, torsion leads to a rapidly decaying but extremely strong repelling force which corresponds to the conservation of half-integer spin \cite{Poplawski2013Spin}. This force \LM{would prohibit} the formation of physical singularities. In the case of cosmology, this leads to so-called ``big bounce'' solutions to the big bang where the point singularity is replaced by either a single or a periodically returning bounce \cite{Poplawski2012, Cubero2019}. The expansion behaviour of the universe, and thus cosmological evolution, is also modified \cite{Piani:2022gon,Piani:2023aof}. \\
As an interesting aside, it was found that the field equations of EC gravity allow for close analogies to electrodynamics (see e.g. \cite{Hehl:1999bt, Hehl:2012pi, Mistretta:2023rnl}) and the $\theta$-term from Yang--Mills theory \cite{Chatzistavrakidis_2020, Chatzistavrakidis_2022}. \\
Allowing non-zero torsion leads to additional antisymmetric terms in the connection coefficients and the field equations of EC gravity \cite{HehlReview76, Poplawski2013, Diether2020}. The antisymmetric terms in the energy-momentum tensor correspond to torsion effects. This is also found when taking the weak-field limit of EC theory \cite{Maitra:2024uhu}. The historical review \cite{HehlReview76} and the contemporary review \cite{Diether2020} argue that this allows for the intrinsic spin of fermionic particles to couple to torsion. This would lead to a theory of gravity that distinguishes between bosonic and fermionic matter. The coupling of fermions and bosons to gravity via torsion (and non-metricity) was also studied in the more general case for metric-affine gravity by \cite{Rigouzzo:2022yan,Rigouzzo:2023sbb}. \\
The last important feature of EC gravity relevant to this work is that torsion vanishes outside of matter sources. The field equations then dictate that torsion does not propagate through vacuum. As a result, EC gravity coincides completely with general relativity in the absence of matter that couples to torsion. Tests of GR in vacuum therefore cannot distinguish between GR and Einstein--Cartan gravity. When considering EC gravity in matter distributions with angular momentum however, there are several stark differences. \\

Note however that torsion-induced effects are still considered non-standard and even speculative by some. While it should be clear that the formulation of EC theory requires less assumptions on the connection, working with general affine connections is more complicated than working with the Levi--Civita connection. Resulting effects like big bounce solutions and the adapted cosmological evolution are not mainstream. \\
The main caveat about torsion effects is the length scale they are expected to occur on: If such effects arise from the intrinsic spin of particles, their typical scale is around $\sim 10^{-26} cm$ \cite{Trautman:2006fp, Boos:2016cey}. Historically, the small scale prompted scientists to discard the idea of torsion effects, even before the concept of intrinsic spin was established. As a result, the understanding of torsion effects in theories of gravity is limited and leaves many areas for both investigation and speculation. \LM{Furthermore, the nature of the coupling between matter and spacetime torsion is also important. For example, non-minimally coupled matter might produce large effects even if only microscopic spin is taken into account, but this type of coupling is still essentially unconstrained by observations (see e.g. \cite{Freidel:2005sn,Alexandrov:2008iy}).} It is therefore of utmost importance to identify possible areas where torsion might play a significant role and to devise experimental tests to verify or dispute its existence. \\ 

To this date, Einstein--Cartan gravity has been studied in a variety of scenarios. Most studies consider the cosmological evolution \cite{Huang:2015zma,deBerredo-Peixoto:2009yvf,Ribas:2009yg,Williams:2011jm} or inflation scenarios \cite{Shaposhnikov:2020gts,Piani:2022gon,Piani:2023aof,GarciadeAndrade:2022fik,He:2023vlj,Boehmer:2005sw,DiMarco:2023ncs}. Lately, there has been a growing interest to consider the effect of EC gravity on the gravitational wave signals of binary compact objects \cite{Kim:1998jv,Battista:2021rlh,Battista:2022hmv,Elizalde:2022vvc,DeFalco:2024ojf}. \\
Some authors also proposed possible experimental tests of EC gravity. One study \cite{ArderucioCosta:2023ooz} proposes to study beams of neutrons in the laboratory and measure the torsion-induced change in their spin polarisation angle. Another study \cite{Capolupo:2023igw} considered the effects of spin-induced torsion on neutrino oscillation. The authors of \cite{Mao:2006bb} proposed to use a satellite in Earth orbit to search for torsion effects, although this specific proposal was debated (see e.g. \cite{Flanagan:2007dc}). These proposals open interesting ways of studying torsion experimentally through its effects on known physical phenomena. \\
One under-explored option to probe EC gravity are neutron stars. They have only been considered by a \LM{small number of studies: \cite{Bohmer:2017dqs} derives modified Tolman-Oppenheimer-Volkoff (TOV) equations and uses them to investigate the Buchdahl limit. \cite{Nussupbekov:2020rjg} uses EC gravity as one possible source of corrections to mass-radius relations in spherically symmetric compact objects using simplified equations of state for the nuclear matter. Finally, \cite{Katkar:2023ugy} formulates models for self-gravitating Weyssenhoff fluid spheres in EC gravity. Our work expands on these previous works by using the TOV equations in EC theory to investigate, in depth, the change of neutron star structure, modelled using a Weyssenhoff fluid. \\
The apparent lack of research on EC gravity in neutron stars} is despite of neutron stars being intensively researched in the wider astrophysical community to probe general relativity \cite{Berti:2015itd,Psaltis:2008bb}, modified theories of gravity \cite{Olmo:2019flu,Manchester:2015mda} and even dark matter \cite{Kouvaris:2010vv,Liebling:2012fv,Bertone:2007ae,Zurek:2013wia,Kouvaris:2013awa,Diedrichs:2023trk,Jockel:2023rrm}. \\

Neutron stars (NSs) are dense and compact remnants of heavy stars. Their high densities make them excellent laboratories for probing gravitation and nuclear physics under extreme conditions. NSs are characterized using the nuclear matter equation of state (EOS). The EOS describes the relation between pressure and energy density of the matter found inside NSs. The total mass, radius and density distribution of neutron stars can be computed from the Tolman-Oppenheimer-Volkoff (TOV) equations \cite{Tolman:1939jz,Oppenheimer:1939ne}. The TOV equations depend on the underlying theory of gravity and on the chosen EOS. Different models of EOS are able to support neutron stars with different radii and total masses. A significant constraint on the EOS is the ability to produce NSs with masses larger than two solar masses, $2\,M_\odot$. The most massive NS known to date is PSR J0952\ensuremath{-}0607 with a mass of $M=2.35^{+0.17}_{-0.17}\,M_\odot$ \cite{Romani:2022jhd}. Such large NS masses require stiff EOS, where the nuclear matter is difficult to compress and the energy density rises sharply with increasing pressure. Other constraints include the measurements of the pulsars PSR J0030\ensuremath{+}0451 ($M=1.34^{+0.15}_{-0.16}\,M_\odot$, $R=12.71^{+1.14}_{-1.19}\,km$) \cite{Vinciguerra:2023qxq} and J0740\ensuremath{+}6620 ($M=2.072^{+0.067}_{-0.066}\,M_\odot$, $R=12.39^{+1.30}_{-0.98}\,km$) \cite{Riley:2021pdl} by the NICER and XMM-Newton telescopes. They also favor a stiff EOS. In contrast, the gravitational wave event GW170817 \cite{Abbott:2018exr,Abbott:2018wiz} favors soft EOS which produce smaller NSs that are more compact and more difficult to tidally disrupt. \\

In this paper, we study neutron stars in Einstein--Cartan gravity. Our aim is to investigate the general effects that the presence of torsion has on NSs and the order of magnitude that torsion effects can have on the NS properties. We discuss the derivation of the adapted TOV equations and the impact and interpretation of the newly arising torsion terms. As possible sources for torsion, we consider mirophysical spin and macroscopic angular momentum. To model the microphysical contributions, we consider possible models of interaction between matter and torsion using spin-fluids. \LM{Throughout we assume that matter is coupled minimally to gravity.} We then derive different relations to model torsion effects induced by macroscopic rotation. We provide some order of magnitude estimations on the expected impact of torsion on various astronomical objects, including neutron stars. \\
We find that the presence of torsion tends to increase the NS density and decrease the radius and gravitational mass. We also find a critical density, above which no stable configuration can exist. Realistic models for microphysical spin lead to torsion effects that have no relevant influence on the neutron star structure. Rotation-induced torsion effects however, can decrease the radius by up to $900\,m$, which is comparable to the increase due to centrifugal forces. Depending on which effect dominates, this leads to a torsion-induced spin-up or spin-down of the neutron star. \\

The paper is structured as follows. In section \ref{sec:theory}, we briefly highlight the main ideas of EC theory and introduce common concepts and notation. In section \ref{sec:spherical-solutions}, we adapt the TOV equations for static neutron stars to the new setting with torsion terms and spin fluids. We also estimate the expected contribution of torsion effects induced by microphysical spin to the NS structure. In section \ref{sec:uniformly-rotating-neutron-stars}, we derive equations to estimate the contributions of torsion effects due to macroscopic rotation. In section \ref{sec:Results}, we investigate the change in neutron star structure with and without rotation-induced torsion. We also discuss possible scenarios where torsion might lead to unstable neutron stars. \\
If not specified otherwise, we use units where $c = G = M_\odot = 1$ throughout this work. \LM{We also assume a $(-,+,+,+)$ metric signature and employ the Einstein summation convention for tensors where Greek indices run over spacetime and Latin indices only over space. In covariant derivatives, we perform summation over the second lower index of the connection coefficients.}

\section{Einstein-Cartan Theory \label{sec:theory}} 
\subsection{Introducing Torsion into Gravity}
\label{subsec:IntroEC}

Einstein--Cartan gravity is a theory of gravity that allows for two geometric properties of affine connections to be non-vanishing: \LM{curvature} and torsion. Lifting the constraint of vanishing torsion, which is standard in literature, allows for a more general class of connections and thus different spacetime-dynamics. Torsion is usually discarded from Einstein gravity a priori by imposing the symmetry of the connection coefficients (Christoffel symbols) $\Gamma_{\mu\nu}^\alpha$ in the lower two indices. EC gravity does not impose this condition and thus we define the \textbf{torsion tensor} as the antisymmetric part of the connection coefficients:
\begin{align}
    \tensor{\tau}{_\mu_\nu^\alpha} := \Gamma_{[\mu\nu]}^\alpha = \frac{1}{2} \left(\Gamma_{\mu\nu}^\alpha - \Gamma_{\nu\mu}^\alpha \right) \ . \label{sec:theory:definition-torsion}
\end{align}
In general the \textbf{connection coefficients} $\Gamma_{\mu\nu}^\alpha$ define an affine connection. We can split them as
\begin{align}
    \Gamma_{\mu\nu}^\alpha := \mathring{\Gamma}_{\mu\nu}^\alpha - \tensor{K}{_\mu_\nu^\alpha} \: , \label{sec:theory:full-connection}
\end{align}
where $\mathring{\Gamma}_{\mu\nu}^\alpha$ are the connection coefficients of the unique metric-compatible connection with vanishing torsion, the Levi--Civita connection. $\tensor{K}{_\mu_\nu^\alpha}$ denote the components of the \textbf{contorsion tensor}
\begin{align}
    \tensor{K}{_\mu_\nu^\alpha} := - \tensor{\tau}{_\mu_\nu^\alpha} + \tensor{\tau}{_\nu^\alpha_\mu} - \tensor{\tau}{^\alpha_\mu_\nu} \LMtwo{= - \tensor{K}{_\mu^\alpha_\nu}} \: . \label{sec:theory:definition-cotorsion}
\end{align}
The contorsion tensor is a measure of the failure of the connection to be torsion-free. The covariant derivative $\nabla_\mu$ is now defined relative to the affine connection (\ref{sec:theory:full-connection}), and thus the \textbf{Riemann tensor} takes the familiar form of (also see \cite{Medina:2018rnl} for more details)
\begin{align}
    \tensor{R}{_\mu_\nu_\alpha^\beta} := \partial_\mu \Gamma_{\nu\alpha}^\beta - \partial_\nu \Gamma_{\mu\alpha}^\beta + \Gamma_{\mu\lambda}^\beta \Gamma_{\nu\alpha}^\lambda - \Gamma_{\nu\lambda}^\beta \Gamma_{\mu\alpha}^\lambda \ . \label{sec:theory:riemann-tensor}
\end{align}
Note that, unlike in usual GR, \eqref{sec:theory:riemann-tensor} includes contributions due to the contorsion tensor $\tensor{K}{_\mu_\nu^\alpha}$, i.e. due to non-vanishing torsion. From the full Riemann tensor one can then define the \textbf{Ricci tensor} as
\begin{align}
    \tensor{R}{_\mu_\nu} := \tensor{R}{_\alpha_\mu_\nu^\alpha} \ . \label{sec:theory:ricci-tensor}
\end{align}
Again, while the form is familiar, the affine connection now includes torsion terms. The curvature scalar $R$ can be obtained from the contraction $R := \tensor{R}{_\mu^\mu} =\tensor{g}{^\mu^\nu}  \tensor{R}{_\mu_\nu}$.\\

This brings us in a suitable situation to define the Lagrangian of EC gravity. First note that to define an affine connection we use two ingredients: A metric $g$, corresponding to a unique $g$-compatible and torsion-free Levi--Civita connection. We also need a contorsion tensor $K$, which describes the failure of a general affine connection to be torsion-free. Thus the curvature scalar $R$, which appears in the Lagrangian of Einstein--Hilbert gravity, can now be seen as a functional $R(g, K)$ of the metric and the contorsion. With $\LL_m$ denoting an arbitrary \textbf{source term}, we then define the action of EC \LM{gravity as \cite{Hehl:1999bt,Medina:2018rnl,Bohmer:2017dqs}}
\begin{align}
    S_{EC}[\tensor{g}{_\mu_\nu}, \tensor{K}{_\mu_\nu^\alpha}] := \int \sqrt{-g} \left(  \frac{1}{2 \kappa} R + \LL_m \right) dx^4 \ . \label{sec:theory:EC-action}
\end{align}
Here $\kappa = 8\pi G/c^4$ is a constant and $g$ denotes the determinant of the metric. Note that the action has the exact same form as that of Einstein--Hilbert gravity, only that now one does not impose vanishing torsion and thus needs to treat $\tensor{K}{_\mu_\nu^\alpha}$ as an independent field.

Before varying the action to obtain the field equations of EC, we define the \textbf{modified torsion tensor}
\begin{align}
    \tensor{M}{_\mu_\nu^\alpha} := \tensor{\tau}{_\mu_\nu^\alpha} +  \delta^\alpha_\mu \tensor{\tau}{_\nu_\lambda^\lambda} - \delta^\alpha_\nu \tensor{\tau}{_\mu_\lambda^\lambda} \CJ{= -\tensor{M}{_\nu_\mu^\alpha}} \: , \label{sec:theory:definition-modified-torsion}
\end{align}
as well as the two tensors $\tensor{T}{_\mu_\nu}$ and $\tensor{\Theta}{_\mu_\nu^\alpha}$ which are defined as the sources of the theory:
\begin{align}
    \tensor{T}{_\mu_\nu}\phantom{_\alpha} :=& - \frac{2}{\sqrt{-g}}\ddelta{\LL_m}{g^{\mu\nu}} \: , \label{sec:theory:definition-EM-tensor} \\
    \tensor{\Theta}{_\mu_\nu^\alpha} :=& - \frac{1}{\sqrt{-g}} \ddelta{\LL_m}{\CJ{\tensor{K}{_\alpha^\mu^\nu}}} \: . \label{sec:theory:definition-EM-tensor-spin}
\end{align}
The equations of motion are then obtained by varying \eqref{sec:theory:EC-action} with respect to the metric $\delta \tensor{g}{^\mu^\nu}$ and the contorsion $\delta \tensor{K}{^\mu^\nu_\alpha}$ respectively \LM{(see \cite{Medina:2018rnl}, section III. A)}:
\begin{subequations} \label{sec:theory:EC-fulll-EOM}
\begin{align}
    \kappa \, \tensor{T}{_\mu_\nu}\phantom{^\alpha} &= G_{\mu\nu} + \nablastar_\lambda (- \tensor{M}{_\mu_\nu^\lambda} + \tensor{M}{_\nu^\lambda_\mu} - \tensor{M}{^\lambda_\mu_\nu}) \: , \label{sec:theory:EC-full-EOM-eq1} \\
    \kappa \, \tensor{\Theta}{_\mu_\nu^\alpha} &= \tensor{M}{_\mu_\nu^\alpha} \: , \label{sec:theory:EC-full-EOM-eq2}
\end{align}
\end{subequations}
where we have defined the $\star$-derivative as \LM{a shorthand notation} $\tensor{\nablastar}{_\mu} := \tensor{\nabla}{_\mu} + 2 \, \tensor{\tau}{_\mu_\alpha^\alpha}$. Looking at \eqref{sec:theory:EC-full-EOM-eq1} and \eqref{sec:theory:EC-full-EOM-eq2} one can interpret $\tensor{T}{_\mu_\nu}$ as the usual energy-momentum-density of a fluid, now including terms due to torsion, and $\tensor{\Theta}{_\mu_\nu^\alpha}$ as describing the fluid properties coupling to torsion. Accordingly we will refer to $\tensor{T}{_\mu_\nu}$ as the \textbf{energy-momentum tensor} and to $\tensor{\Theta}{_\mu_\nu^\alpha}$ as the \textbf{spin-energy tensor}. Together, the equations \eqref{sec:theory:EC-fulll-EOM} can be interpreted to model the coupling of spacetime to energy (\eqref{sec:theory:EC-full-EOM-eq1}) and to angular momentum (\eqref{sec:theory:EC-full-EOM-eq2}), respectively. \\
\newline
The equations of motion \eqref{sec:theory:EC-fulll-EOM} can be combined into a single equation by inserting \eqref{sec:theory:EC-full-EOM-eq2} into \eqref{sec:theory:EC-full-EOM-eq1}:
\begin{align}
    \tensor{\mathring{G}}{_\mu_\nu} &= \kappa \, \tensor{T}{_\mu_\nu} + \kappa^2 \Big[ -4 \tensor{\Theta}{_\mu_\lambda^{[\alpha}} \tensor{\Theta}{_\nu_\alpha^{\lambda]}} - 2 \tensor{\Theta}{_\mu_\lambda_\alpha} \tensor{\Theta}{_\nu^\lambda^\alpha} \nonumber \\
    &+ \tensor{\Theta}{_\alpha_\lambda_\mu} \tensor{\Theta}{^\alpha^\lambda_\nu} + \frac{1}{2} \tensor{g}{_\mu_\nu} \big( 4 \tensor{\Theta}{_\lambda^\beta_{[\alpha}} \tensor{\Theta}{^\lambda^\alpha_{\beta]}} + \tensor{\Theta}{_\alpha_\lambda_\beta} \tensor{\Theta}{^\alpha^\lambda^\beta}\big) \Big].\label{sec:theory:full-einstein-cartan-equation}
\end{align}
Here $\tensor{\mathring{G}}{_\mu_\nu} = \tensor{\mathring{R}}{_\mu_\nu} - \frac{1}{2} \mathring{R} \, \tensor{g}{_\mu_\nu}$ is the \textbf{Einstein tensor} associated to the Levi--Civita connection defined by $\tensor{g}{_\mu_\nu}$. We have collected all terms related to the source on the right-hand-side. In this form (\eqref{sec:theory:full-einstein-cartan-equation}), torsion can be interpreted as a correction to the energy-momentum tensor. Note that – due to the minimal coupling assumption – all terms containing contributions due to torsion are proportional to $\kappa^2 = 64\pi^2 G^2/c^8$. As a result one should expect torsion effects to be small compared to curvature effects. We will later explore in which situations one can expect significant contributions due to torsion. \\
\LM{We also note a third approach to EC theory. In \cite{Kibble:1961ba} it is shown that torsion can be "integrated out" and that EC theory can be recast into an equivalent metric theory without torsion but with an additional fermion-four interaction term. This term is suppressed by the coupling constant $\kappa$. When computing the variation of Eq. (7.1) in \cite{Kibble:1961ba}, one obtains the $\kappa^2$-term in our \eqref{sec:theory:full-einstein-cartan-equation}. This underlines the equivalence of these different approaches of EC theory.}

\subsection{Sources of Torsion: Spin-fluids}
\label{subsec:SpinFluids}

In Einstein-Cartan theory, particles with intrinsic spin and fluids composed of them couple to gravity through torsion effects \cite{HehlReview76,Diether2020}. Rotating macroscopic bodies might be another possible source of torsion \cite{Mao:2006bb,Bohmer:2017dqs}. We will discuss this in section \ref{sec:uniformly-rotating-neutron-stars}. Thus all rotational properties of a fluid might serve as the source of the spin-energy tensor $\tensor{\Theta}{_\mu_\nu^\alpha}$. The spin-properties of a particle with intrinsic spin are described using the antisymmetric \textbf{spin-density tensor} \cite{Frenkel:1926nat,Bohmer:2017dqs}. It has the units of an angular momentum density:
\begin{align}
    \tensor{s}{_\mu_\nu} = - \tensor{s}{_\nu_\mu}. \label{sec:theory:eq:anti-symmetric-spin-density-definition}
\end{align}
The components $(s_{23}, s_{31}, s_{12})$ correspond to the magnetic moment of the spinning particle(s) (i.e. the spin-vector) in the rest-frame and the components $(s_{01}, s_{02}, s_{03})$ describe the electrical moment of the particle(s) \cite{Frenkel:1926nat,Obukhov:1987}.

In this work we consider a macroscopic fluid consisting of particles with half-integer quantum-mechanical spin. Such a fluid is commonly called a \textbf{Weyssenhoff fluid}. On a macroscopic scale a Weyssenhoff fluid is characterized by the spin properties of its constituent particles. In this case $\tensor{s}{_\mu_\nu}$ serves as the only source of the spin-energy tensor $\tensor{\Theta}{_\mu_\nu^\alpha}$ via
\begin{align}
    \tensor{\Theta}{_\mu_\nu^\alpha} = \tensor{s}{_\mu_\nu} \tensor{u}{^\alpha} \ ,
    \label{sec:theory:eq:spin-density-tensor-spinfluid}
\end{align}
where $\tensor{u}{^\alpha}$ denotes the four-velocity of the fluid. \\
\newline
The spin-density tensor satisfies the \textbf{Frenkel condition} \cite{Frenkel:1926nat}, which states that in the rest-frame of a spinning particle, its electrical moment vanishes. This condition can be formulated as the covariant expression \cite{Frenkel:1926nat,Bohmer:2017dqs}
\begin{align}\label{sec:theory:eq:FrenkelCondition}
    \tensor{s}{_\mu_\nu} \tensor{u}{^\nu} = 0 \ .
\end{align}
In any physical scenario on the scale of compact astrophysical objects, it is reasonable to assume that the spin contribution of each individual particle in the spin fluid can be described using space-averaged quantities. This allows for a macroscopic description of the fluid using the \textbf{spin-density scalar} \cite{Bohmer:2017dqs,Obukhov:1987}
\begin{align}\label{sec:theory:def-spin-energy-density}
    s^2 = \frac{1}{2} \tensor{s}{_\mu_\nu} \tensor{s}{^\mu^\nu} \ .
\end{align}
The spin-density scalar of a fluid of half-integer spin (fermions) with number density $n$ and no spin polarization (i.e. randomly oriented spins) is given by \cite{Nurgalev:1983vc,Gasperini:1986mv,Bohmer:2017dqs,Atazadeh:2014ysa}
\begin{align}
    s^2 = \frac{1}{8} \left( \hbar n \right)^2 \ . \label{sec:theory:eq:spin-fluid-eos-fermions}
\end{align}
It should be noted that, to our knowledge, no detailed derivation of the above formula exists in literature. The formula is first used in \cite{HehlReview76} and subsequently claimed to be true in \cite{Nurgalev:1983vc}. We therefore advise caution when using the above expression. Note that this form of the spin-density scalar does not affect the conclusions of this work. \\
Equation (\ref{sec:theory:eq:spin-fluid-eos-fermions}) takes the role of an equation of state for the spin-density scalar. Apart from the spin-contributions of the individual particles to the fluid, the bulk properties of the fluid must also be considered. All properties of the Weyssenhoff fluid can be modeled using a modified version of the energy-momentum tensor of a perfect fluid, which respects the contributions of intrinsic spin to the macroscopic fluid-properties. This leads to the following energy-momentum tensor (see \cite{Ray:1982qr}\LM{, also section III. C of \cite{Medina:2018rnl} for a derivation)}:
\begin{align}
    \tensor{T}{^\mu^\nu} &= \left( e + P \right) \tensor{u}{^\mu} \tensor{u}{^\nu} + P \tensor{g}{^\mu^\nu} + \tensor{\nablastar}{_\lambda} \left( \tensor{s}{^\nu^\lambda} \tensor{u}{^\mu} - \tensor{s}{^\lambda^\mu} \tensor{u}{^\nu} \right) \nonumber \\
    &+ \tensor{u}{^\lambda} \tensor{u}{^{\nu}} \tensor{\mathring{\nabla}}{_\alpha} \left( \tensor{u}{^\alpha} \tensor{s}{^{\mu}_\lambda} \right) + \tensor{u}{^\lambda} \tensor{u}{^{\mu}} \tensor{\mathring{\nabla}}{_\alpha} \left( \tensor{u}{^\alpha} \tensor{s}{^{\nu}_\lambda} \right) \ .
    \label{sec:theory:eq:Energy-momentum-tensor-perfect-fluid-w-spin-contributions}
\end{align}
$\tensor{T}{^\mu^\nu}$ consists of a perfect fluid part and a part accounting for the spin contributions to the fluid. The last two terms will vanish due to the Frenkel conditions \eqref{sec:theory:eq:FrenkelCondition} as soon as we assume vanishing acceleration \cite{Obukhov:1987, Bohmer:2017dqs}. This is the case e.g. for a static fluid. \\
\newline
Apart from relation (\ref{sec:theory:eq:spin-fluid-eos-fermions}), there are also simpler models for the spin fluid like imposing a constant spin-density scalar $s^2 = const.$. Another option is to use a phenomenological model where the spin-density scalar is proportional to some power of the pressure \cite{Bohmer:2017dqs}:
\begin{align}
    s^2 = \beta P^\gamma \: , \label{sec:theory:eq:Polytropic-sin-density-ansatz}
\end{align}
where $\beta$ and $\gamma$ are scalar constants.
\section{Spherically symmetric static solutions \label{sec:spherical-solutions}} 

\subsection{Equations of Motion} \label{subsec:background-spherical}

We solve the Einstein-Cartan equations for a static (zero spatial velocity and zero acceleration) spin-fluid obeying the Frenkel condition (\ref{sec:theory:eq:FrenkelCondition}) by inserting the ansatz for the spin-density tensor (\ref{sec:theory:eq:spin-density-tensor-spinfluid}) and the energy-momentum tensor (\ref{sec:theory:eq:Energy-momentum-tensor-perfect-fluid-w-spin-contributions}), discussed in the previous section \ref{subsec:SpinFluids}, into equations (\ref{sec:theory:full-einstein-cartan-equation}) to obtain
\begin{align}
    \tensor{\mathring{G}}{_\mu_\nu} &= \kappa \left\{ \left( e_{\mathrm{eff}} + P_{\mathrm{eff}} \right) \tensor{u}{_\mu} \tensor{u}{_\nu} + P_{\mathrm{eff}} \tensor{g}{_\mu_\nu} \right\} \: , \label{sec:spherical-solution:full-einstein-cartan-equation-analogy-GR}
\end{align}
where we defined the \textbf{effective pressure} and \textbf{effective energy-density} of the spin-fluid
\begin{align}
\label{eq:effective_quantities}
    e_{\mathrm{eff}} := e - \kappa s^2 \:\: , \:\: P_{\mathrm{eff}} := P - \kappa s^2 \: .
\end{align}
This leads to the Einstein-Cartan equations (\ref{sec:spherical-solution:full-einstein-cartan-equation-analogy-GR}) having the same form as the regular Einstein equations, known from GR, coupled to a perfect fluid. \\
There are now two ways to go forward: Either one uses the effective quantities and solves the EC equations with respect to these quantities. Or one formulates the solution using the pressure $P$ and the spin-density $s^2$. We opt for the second option since the first option obscures the phenomenology of the spin-contributions. \\
\newline
We assume a static spacetime with spherical symmetry and chose an ansatz with metric functions $\alpha(r)$ and $a(r)$ that depend solely on the radius:
\begin{align}
    d s^2 = - \alpha^2 dt^2 + a^2 dr^2 + r^2 d \theta^2 + r^2 \sin(\theta)^2 d \varphi^2 \: .
\end{align}
We further use the fact that $\tensor{\mathring{\nabla}}{_\mu} \tensor{\mathring{G}}{^\mu^\nu} = 0$ and apply the covariant derivative associated to the Levi-Civita connection $\tensor{\mathring{\nabla}}{_\mu}$ to the right hand side of equation (\ref{sec:spherical-solution:full-einstein-cartan-equation-analogy-GR}). This leads to a conversation equation for effective energy-momentum. The four-velocity for a static fluid is given by $u^\mu = (1/\alpha, 0, 0, 0)$ and $u_\mu = (-\alpha, 0, 0, 0)$ and is normalized as $u_\mu u^\mu = -1$. Following these steps one obtains the \textbf{Tolmann-Oppenheimer-Volkoff (TOV)} equations in EC gravity for a Weyssenhoff spin-fluid:
\begin{subequations}
\label{eq:TOV_collection}
\begin{align}
	a' = \frac{da}{dr} &=  \frac{a}{2} \, \left[ \frac{(1-a^2)}{r} + \kappa r a^2 \left( e - \kappa s^2 \right) \; \right] \: , \label{eq:TOV_collection-a-eq} \\
	\alpha' = \frac{d \alpha}{dr} &= \frac{\alpha}{2} \left[ \frac{(a^2 -1)}{r} + \kappa r a^2 \left( P - \kappa s^2 \right) \right] \: , \label{eq:TOV_collection-alpha-eq} \\
	P' = \frac{dP}{dr} &= - (e + P - 2 \kappa s^2) \frac{\alpha'}{\alpha\,} + \kappa \frac{d(s^2)}{dr} \: . \label{eq:TOV_collection-P-eq}
\end{align}
\end{subequations}
The system of equations is closed by an equation of state $P(e,\rho)$ relating the fluid pressure $P$ to the energy density $e$ and the restmass density $\rho$. In addition, an ``equation of state'' for the spin-density $s^2$ is needed. In the following we will investigate three options of relating $s^2$ to the other hydrodynamic quantities
\begin{itemize}
    \item[a)] \textbf{$s^2 = const. = s_0^2$:} This leads to the simplest modification of the TOV equations by introducing a constant offset to both energy density and pressure. For $s^2 = s_0^2 = const.$, equation (\ref{eq:TOV_collection-P-eq}) takes the following form:
    \begin{align}
        P' &= - (e+P - 2 \kappa s_0^2) \frac{\alpha'}{\alpha\,} \: . \label{eq:constant-spin-density-model}
    \end{align}

    \item[b)] \textbf{$s^2 = \beta P^\gamma$:} Mimicking a polytropic equation of state, this ansatz can be understood as a parallel to using a polytropic equation $P = k \rho^\gamma$ for the restmass density. Using this ''polytropic'' ansatz for $s^2$,  equation (\ref{eq:TOV_collection-P-eq}) becomes
    \begin{align}
        P' &= - \frac{(e+P - 2 \kappa \beta P^\gamma)}{(1 - \kappa \beta\gamma P^{\gamma -1})}\, \frac{\alpha'}{\alpha\,} \: . \label{eq:polytropic-spin-density-model}
    \end{align}
    The parameter $\beta$ can be understood as the strength of the spin-coupling to spacetime. We chose $\beta$ of the order $\mathcal{O}(10^0-10^2)$ to match the parameter range used in \cite{Bohmer:2017dqs}.

    \item[c)] \textbf{$s^2 = \frac{1}{8} (\hbar n)^2$:} This ansatz describes a fermion fluid with number density $n$ and no spin polarization, see \eqref{sec:theory:eq:spin-fluid-eos-fermions}. Equation (\ref{eq:TOV_collection-P-eq}) changes to
    \begin{subequations}
    \label{eq:realistic-spin-density-model}
    \begin{align}
    	&P' = \kappa \frac{\hbar^2}{4} n \, n' - \left(e+P - \frac{\kappa}{4}   (\hbar n)^2\right) \frac{\alpha'}{\alpha\,} \: , \\
     &= \left(e+P - \frac{\kappa}{4} (\hbar n)^2\right) \left(1 - \kappa \frac{\hbar^2}{4} n \frac{\partial n}{\partial P} \right)^{-1} \frac{\alpha'}{\alpha\,} \: .
    \end{align}
    \end{subequations}
    For the second equality we used that, if $n=n(P)$, one can use $dn/dr=\partial n/\partial P \times dP/dr$ and solve for $P'$.
\end{itemize}

At last, we define some relevant global quantities of neutron stars. The radius $R$ is defined as the point where the pressure $P$ becomes zero (in numerical practice, the value is chosen as a small number $\approx 10^{-15}$). Because in vacuum, Einstein--Cartan gravity is equivalent to general relativity, the Birkhoff--Jebsen\footnote{\LM{The commonly known ``Birkhoff theorem'' should also include the name of Norwegian physicist J\o rg Tofte Jebsen, because he discovered the theorem two years before Birkhoff. Unfortunately, Jebesns work went largely unnoticed. See \cite{VojeJohansen:2005nd} for a short historical summary.}} theorem also holds and the outer solution is given by the Schwarzschild metric. The \textbf{gravitational mass} can then be computed using the metric functions as
\begin{align}
    M_{grav} := \lim_{r\rightarrow \infty} \frac{r}{2} \left( 1 - \frac{1}{(a(r))^2} \right) \: .
\end{align}
In practice, the mass can be extracted at $r=R$ since the gravitational mass will not change for larger distances. We define the \textbf{rest mass} $M_{rest}$ as the spatial integral over the conserved four-current $J^\mu = \rho u^\mu$ of neutron star matter:
\begin{align}
    M_{rest} := \int \sqrt{-g}\, J_\mu g^{\mu t} dx^3 = 4\pi \int_0^R a \rho r^2 dr \: . 
\end{align}

\subsection{Analytical Results} \label{subsec:analytical-results}

In this section, we derive an analytic estimate for the scale at which the torsion effects derived so far become relevant. In particular we investigate if typical neutron stars provide the necessary environment to probe torsion. We focus hereby on the torsion effects sourced by microphysical spin. The effect of macroscopic rotation will be discussed in section \ref{sec:uniformly-rotating-neutron-stars}. \\

Equation (\ref{sec:theory:eq:spin-fluid-eos-fermions}) describes a fermion fluid with no spin polarization with a number density $n$.  The equation can also be rewritten using the fermion mass $m_\mathrm{f}$ and the restmass density $\rho$:
\begin{align}
\label{eq:s2_fermion_ansatz}
    s^2 = \frac{1}{8} \left( \hbar \frac{\rho}{m_\mathrm{f}}\right)^2 \: .
\end{align}
The goal is to find out if and at what densities such a system produces significant contributions to the effective energy density \LM{(in SI units)} $e_\mathrm{eff} = e -  \kappa \LM{c^2} s^2$ (see \eqref{eq:effective_quantities}), which has contributions from the energy density $e$ and the spin-density scalar $s^2$. We derive an expression to find out where the spin effects have a certain relative magnitude $\eta := \kappa \LM{c^2} s^2/e$ compared to the energy density. Locally, the energy density can be written as a restmass density $e = \LM{c^2} \rho$. \LM{By setting $\kappa c^2 s^2$} equal to $\eta \times e$ one thus obtains:
\begin{align}
     \rho = \eta \frac{8 m_\mathrm{f}^2}{\hbar^2 \kappa} \: .
\end{align}
For a given fermion with mass $m_\mathrm{f}$ this is the density at which the spin-effects have a relative contribution of $\eta$. As an example, we consider the case where the spin density of neutrons contribute $1\%$ ($\eta = 0.01$) of the energy density $e$. We obtain a density of $\rho \simeq 9.718 \cdot 10^{50} kg/m^3$. This is far beyond the typical densities found in neutron stars of just several times the nuclear saturation density $\rho_\mathrm{sat} \approx 2.7 \cdot 10^{17} kg/m^3 = n_{nuc} m_n$ \cite{Lattimer:2021emm}, where $m_n$ is the neutron mass and $n_{nuc}=0.16\,$fm$^{-3}$ is the average nucleon number density. This situation is similar for all fermions known to the standard model. To our understanding, densities in this order of magnitude can only be expected in the early universe and near black hole singularities. The impact of torsion effects in neutron stars due to microphysical spin is therefore negligible. \LM{This result is consistent with earlier works that estimated that torsion effects only become important at very high densities (see e.g. \cite{HehlReview76}).} \\

Note however that for fermions with small masses, the spin contributions relative to the matter density become significantly larger. As a result, very light fermionic particles with masses of around $10^{-9} - 10^{-12}\, eV$ only require a matter density of roughly the nuclear saturation density $\rho_\mathrm{sat}$ for their spin to have significant effects compared to the energy density. If torsion exists and couples minimally to fermions, this opens up the possibility to probe for ultralight fermionic particles beyond the standard model, for example as dark matter candidates, in this mass-range using neutron stars. \LM{Whether this approach is viable, of course, depends on the dark matter model used. But it should be noted that previous bounds on various types of fermionic dark matter (such as sterile neutrinos and self-interacting dark matter) imply particle masses in the $>keV$ range \cite{Arguelles:2023nlh,Tremaine:1979we}.}
\section{Uniformly rotating Neutron Stars \label{sec:uniformly-rotating-neutron-stars}} 
So far we considered microphysical spin as a source for torsion effects. However, the angular momentum of a macroscopic object \LM{– irrespective of the matter that makes it up –} might be another possible source of torsion effects, as argued e.g. by \cite{Mao:2006bb}. \LMtwo{In particular, even objects made up entirely of bosonic matter could source torsion, as long as they carry macroscopic angular momentum. This of course casts some doubt on the proposal, since – so far – only fermions have been shown to couple to torsion \cite{HehlReview76, Kibble:1961ba}. We further discuss this below.} \\

The goal of this section is to estimate the magnitude of such effects using the macroscopic angular momentum of a rotating neutron star. In particular, we investigate how large the torsion effects are compared to the energy-density effects of the neutron star fluid. However, we expect that torsion effects will always be smaller simply by virtue of the quadratic coupling constant $\kappa=8\pi G/c^4$ in the field equations \eqref{sec:theory:full-einstein-cartan-equation}. \\

From a geometric viewpoint, curvature corresponds to the failure of parallel transports to commute. As a result, the masses of single particles and the emergent mass of macroscopic objects all induce curvature by means of that same geometric description. Similarly, torsion describes the failure of parallel transport parallelograms to close. Employing the same geometric reasoning to microscopic spin and macroscopic rotation, one might very well expect torsion effects induced from both. \\

Some authors however, e.g. \cite{HehlReview76}, claim that torsion can only couple to intrinsic spin (i.e. to intrinsic angular momentum). The argument is that, on a sufficiently small scale, all movement due to a macroscopic spin will look like parallel movement in a single direction. This adheres to the gauge-theoretic view on curvature and torsion. There, the symmetric momentum currents are induced by the translation part of the Lorentz group and the antisymmetric spin currents by the rotation part (for more details, see \cite{HehlReview76, Gronwald:1995em}). Thus it makes sense – when inspecting very small open neighbourhoods of particles in a neutron star – to assume that the macroscopic rotation of the object does not induce substantial local effects. \\

Following the gauge-theoretic view on spin-torsion coupling, there would be a ``small'' length scale below which one has to consider the coupling of angular momentum to torsion. However it is unclear what this length scale should be. For example, one might consider small vortices in collections of moving particles (e.g. charged particles gyrating around magnetic field lines). Depending on the length scale, one would have to include the effective angular momentum as a source for torsion or not. An instance of this are so-called ``oscillons'' \cite{Zhang:2021xxa}: Oscillon configuration arise in collections of self-gravitating particles with a periodically rotating vector field associated to them. They carry effective intrinsic spin directly proportional to the number of particles involved but have a macroscopic size. In the case of neutron stars, we could treat them as a macroscopic vortex of particles, which is why their angular momentum might be a potential source of torsion. We argue that this does not necessarily conflict with the gauge-theoretic viewpoint: If we see EC as a gauge theory for the Poincaré group, then one would expect an associated microscopic and macroscopic conserved quantity for both the translation and the rotation part of the group. \\

In the following, we consider the possibility that torsion also couples to macroscopic rotation. The basic idea is to find an expression for the spin-density tensor $\tensor{\Theta}{_\mu_\nu^\alpha} = \tensor{s}{_\mu_\nu} \tensor{u}{^\alpha}$, such that it describes macroscopic rotation instead of intrinsic spin (see \autoref{subsec:SpinFluids}). There are a number of possible approaches, two of which we consider in detail. Because we are only interested in an order of magnitude estimation, we make a number of simplifying assumptions. We assume that the neutron star is
\begin{enumerate}[label=\alph*)]
    \item spherically symmetric (thus there is no $(\theta, \varphi)$-dependencies),
    \item uniformly rotating with a constant angular velocity $\Omega = const.$.
\end{enumerate}
We further consider the spacetime to be flat. This entails errors on the order of a few ten percent due to neglected curvature effects but will simplify our calculations considerably. The accuracy will be sufficient to obtain the correct order of magnitude. \\

For the first approach, we approximate the neutron star as a rotating rigid sphere. We use the \textbf{relativistic angular momentum tensor} $\tensor{L}{^\mu^\nu} = x^\mu p^\nu - x^\nu p^\mu$ \cite{fayngold2008special}\footnote{This is a tensorial generalization of $L= r \times p$}, where $x^\mu$, $p^\mu$ are the four-position and four-momentum respectively. It can be related to the \textbf{four-spin} (Pauli--Lubanski pseudovector) $S^\mu$ \cite{LUBANSKI-1,LUBANSKI-2} such that $S_\mu = \epsilon_{\mu\nu\rho\sigma} L^{\nu\rho} p^\sigma$, where $\epsilon_{\mu\nu\rho\sigma}$ is the totally antisymmetric symbol. In the particle rest frame, the components of $S^\mu$ are related to the angular momentum of the particle via
\begin{align}
    S_\mu = (0, L^{yz}, L^{zx}, L^{xy}) \: . \label{sec:uniformly-rotating-neutron-stars:eq:pauli-lubanski-pseudovector}
\end{align}
Without loss of generality, we assume that the angular momentum is entirely aligned along the $z$-axis. The angular momentum $S_z$ is then equal to the total magnitude of the angular momentum $L$, which we take to be equal to that of a rigid uniformly rotating sphere. Thus
\begin{align}
    S_z = L^{xy} = L = \frac{2}{5} M R^2 \Omega \: , \label{sec:uniformly-rotating-neutron-stars:eq:pauli-lubanski-pseudovector-sphere-angular-momentum}
\end{align}
where $\Omega$ is the angular velocity. Now that we have defined the components of the angular momentum tensor $L^{\mu\nu}$, we define the \textbf{spin-density tensor} (see \eqref{sec:theory:eq:anti-symmetric-spin-density-definition}) as
\begin{align}
    \tensor{s}{^\mu^\nu} := \frac{\tensor{L}{^\mu^\nu}}{V_{NS}} \: , \label{sec:uniformly-rotating-neutron-stars:eq:redefinition-spin-tensor-pointparticle-ns}
\end{align}
where $V_{NS} = \frac{4}{3} \pi R^3$ is the volume of the neutron star. It follows that the spin-density scalar \eqref{sec:theory:def-spin-energy-density} is given by
\begin{align}
    s^2 = \frac{9}{100\pi^2} \frac{M^2}{R^2} \Omega^2 \: . \label{sec:uniformly-rotating-neutron-stars:eq:spin-density-pointparticle-ns}
\end{align}
For a Keplerian rotation velocity of $\Omega^2 = M/R^3$ and typical values for a neutron star of $M=2\,M_\odot$ and $R= 8\,M_\odot$ ($\approx 12.5\,km$), we obtain a value of $s^2 \approx 2.226 \e{-6}$. In comparison to the average density of the star $\Bar{e} = M / V_{NS} \approx 9.325 \e{-4}$, the relative magnitude of the torsion effects to the energy-density effects (see \eqref{eq:effective_quantities}) are expected to be of order $\kappa s^2 / \Bar{e} \approx 6\,\%$. This suggests at most a percent-order correction to the neutron star properties due to torsion effects. However, note that realistic neutron stars rotate at significantly lower rates than the Keplerian limit, which is why the impact of torsion will likely be even smaller. \\
Repeating the same calculation as above for the Sun ($M=1\,M_\odot$, $R=471000\,M_\odot$) leads to $s^2 \approx 3.9 \e{-31}$, $\Bar{e} \approx 2 \e{-18}$ and finally to a relative torsion contribution of $\kappa s^2 / \Bar{e} \approx 5 \e{-10}\,\%$. This suggests that torsion effects due to rotation are negligible for main sequence stars like the Sun. \\

Our second approach is more sophisticated. We assume that the neutron star is a spherical rigidly rotating matter distribution. For that we make use of the relativistic generalization of the angular momentum tensor for extended bodies as discussed by \cite{Dixon:1970zza,Dixon:1970zz,Dixon:1974xoz}. The angular momentum tensor is computed using the integral
\begin{align}
    L^{\alpha\beta} = \oint_{\partial U} \mathcal{M}^{\alpha\beta\gamma} d \Sigma_\gamma \: ,
\end{align}
where $\Sigma_\gamma$ is a timelike unit vector on the 3-dimensional boundary $\partial U$ of the spacetime domain. Here, the tensor $\mathcal{M}^{\alpha\beta\gamma}$ has the units of angular momentum density and is given by
\begin{align}
    \tensor{\mathcal{M}}{^\alpha^\beta^\gamma} := ( \tensor{x}{^\alpha} - x_0^\alpha)\, \tensor{T}{^\beta^\gamma} - (\tensor{x}{^\beta} - x_0^\beta )\, \tensor{T}{^\alpha^\gamma} \: , \label{sec:uniformly-rotating-neutron-stars:eq:spin-density-three-tensor}
\end{align}
where $\tensor{T}{^\mu^\nu}$ is the energy-momentum tensor of the fluid, $x^\mu$ is the four-position vector of a given fluid element, and $x_0^\mu$ describes the rotational axis and is constant. \eqref{sec:uniformly-rotating-neutron-stars:eq:spin-density-three-tensor} has the units of angular momentum density and is the Noether current for rotations in spacetime. It is conserved if the energy-momentum tensor is symmetric, i.e. $\tensor{T}{^\mu^\nu} = \tensor{T}{^\nu^\mu}$. Choosing $U$ such that its boundary is a spacelike surface of constant time, one obtains
\begin{align}
    L^{\alpha \beta} &= \oint_{\partial U} \mathcal{M}^{\alpha \beta t} d \Sigma_t \: . \label{sec:uniformly-rotating-neutron-stars:eq:angular-momentum-tensor}
\end{align}
The components $\mathcal{M}^{\alpha \beta t}$ thus can be interpreted as describing the angular momentum density of an extended matter distribution at a fixed time. \\
We thus define the spin-density tensor \eqref{sec:theory:eq:anti-symmetric-spin-density-definition} as the projection of the angular momentum density \eqref{sec:uniformly-rotating-neutron-stars:eq:spin-density-three-tensor} onto a spacetime slice of constant time:
\begin{align}
    \tensor{s}{^\mu^\nu} := \tensor{\mathcal{M}}{^\mu^\nu^\alpha} \tensor{n}{_\alpha} \: . \label{sec:uniformly-rotating-neutron-stars:eq:redefinition-spin-tensor-extended-body-ns}
\end{align}
Here $n^\mu$ is a dimensionless timelike unit vector which is normal to a surface of constant time. In the special-relativistic context we are considering here, $n^\mu = (1,0,0,0)$ and $n^\mu n_\mu = -1$. \\
We now compute the components $\mathcal{M}^{\alpha \beta t}$. For that, we assume that the neutron star is described by a perfect fluid in flat spacetime, thus:
\begin{align}
    \tensor{T}{^\mu^\nu} &= \left( e + P \right) \tensor{u}{^\mu} \tensor{u}{^\nu} + P \tensor{\eta}{^\mu^\nu} \: . \label{sec:uniformly-rotating-neutron-stars:eq:extended-body-energy-momentum-tensor}
\end{align}
$e$ and $P$ are the energy density and pressure of the fluid respectively. To simplify our calculations, we use a Cartesian coordinate basis expressed in spherical coordinates to compute the tensor components. We do this because the four-position vector $x^\mu$ is not well defined for non-Cartesian coordinate bases. Without loss of generality we assume that the rotational axis of the neutron star is aligned with the $z$-axis. The four-velocity of a given fluid element in an uniformly rotating neutron star is then given by
\begin{align}
    \tensor{u}{^\mu} = \Gamma \left(\, 1,\, -r\Omega \sin(\phi),\, r\Omega \cos(\phi),\, 0 \,\right) \: . \label{sec:uniformly-rotating-neutron-stars:eq:extended-body-four-velocity}
\end{align}
Here $\Omega$ is the angular rotation velocity and we have defined the Lorentz factor $\Gamma := 1/\sqrt{1 - r^2\Omega^2}$. The four-position vector is given by
\begin{align}
    \tensor{x}{^\mu} =& \left(\, t,\, x,\, y,\, z \,\right) \label{sec:uniformly-rotating-neutron-stars:eq:extended-body-four-position} \\
    =& \left(\, t,\, r \cos(\phi)\sin(\theta),\, r\sin(\phi)\sin(\theta),\, r \cos(\theta) \,\right) \: . \nonumber
\end{align}
Accordingly, the axis of rotation $x_0^\mu$ is given by
\begin{align}
    x_0^\mu =& \left(\, t,\, 0,\, 0,\, z \,\right) = \left(\, t,\, 0,\, 0,\, r \cos(\theta) \,\right) \: . \label{sec:uniformly-rotating-neutron-stars:eq:extended-body-four-position-rotation-axis}
\end{align}
Using the above information, we compute the components of the energy-momentum tensor. Subsequently, we compute the components of the angular momentum density tensor (\eqref{sec:uniformly-rotating-neutron-stars:eq:spin-density-three-tensor}). We show the components of $T^{\mu\nu}$ in Appendix \ref{sec:appendix:angular-momentum-tensor-of-extended-body}  \\
The $t$-components of the spin-density tensor can be ``gauged away'' by imposing the condition
\begin{align}
    L_{\mu\nu} P^\nu = 0 \: , \label{sec:uniformly-rotating-neutron-stars:eq:dixon-condition}
\end{align}
which was proposed by Dixon in \cite{Dixon:1970zz}. Here $P^\mu$ is the four-momentum of the macroscopic extended body, defined as
\begin{align}
    P^{\mu} = \oint_{\partial U} T^{\mu \alpha} d \Sigma_\alpha  \: , \label{sec:uniformly-rotating-neutron-stars:eq:extended-body-four-momentum-extended-body}
\end{align}
where $\Sigma$ and $U$ are defined as above. The condition \eqref{sec:uniformly-rotating-neutron-stars:eq:dixon-condition} is thus equivalent to choosing the center of mass frame of the neutron star. Note that this does \textit{not} coincide with the co-rotating frame in general. The Dixon condition \eqref{sec:uniformly-rotating-neutron-stars:eq:dixon-condition} leads to the components $L_{t\mu}$ being zero. When considering the components of \eqref{sec:uniformly-rotating-neutron-stars:eq:angular-momentum-tensor} individually, it is possible to also extend this constraint to the components of $\mathcal{M}^{\alpha \beta t}$ and thus also to the components of $\tensor{s}{^\mu^\nu} = \mathcal{M}^{\mu \nu t} n_t$. Therefore, we can set $\tensor{s}{^t^\nu} = 0$\footnote{Note that when imposing the Dixon condition \eqref{sec:uniformly-rotating-neutron-stars:eq:dixon-condition}, the Frenkel condition \eqref{sec:theory:eq:FrenkelCondition} does not necessarily hold any more since the latter uses the four-velocity of a fluid particle. The former uses the four-momentum of the whole body.}. More information on this argument is provided in Appendix \ref{sec:appendix:details-on-gauging-away-components-of-the-spin-density-tensor}. \\
The remaining non-zero components of the spin-density tensor are
\begin{align}
    \tensor{s}{^x^y} = \tensor{\mathcal{M}}{^x^y^t} &= r^2 \Omega \Gamma^2 (e+P) \sin(\theta) \: .
\end{align}
The spin-density scalar \eqref{sec:theory:def-spin-energy-density} then takes the form
\begin{align}
    s^2 = r^4 \Omega^2 \Gamma^4 (e+P)^2 \sin^2(\theta) \: . \label{sec:uniformly-rotating-neutron-stars:eq:spin-density-extended-body-ns}
\end{align}
Finally it should be noted that all calculations here are performed in the setting of special relativity. We leave a consistent general-relativistic version for future work. \\

\begin{figure}
\centering
    \includegraphics[width=0.495\textwidth]{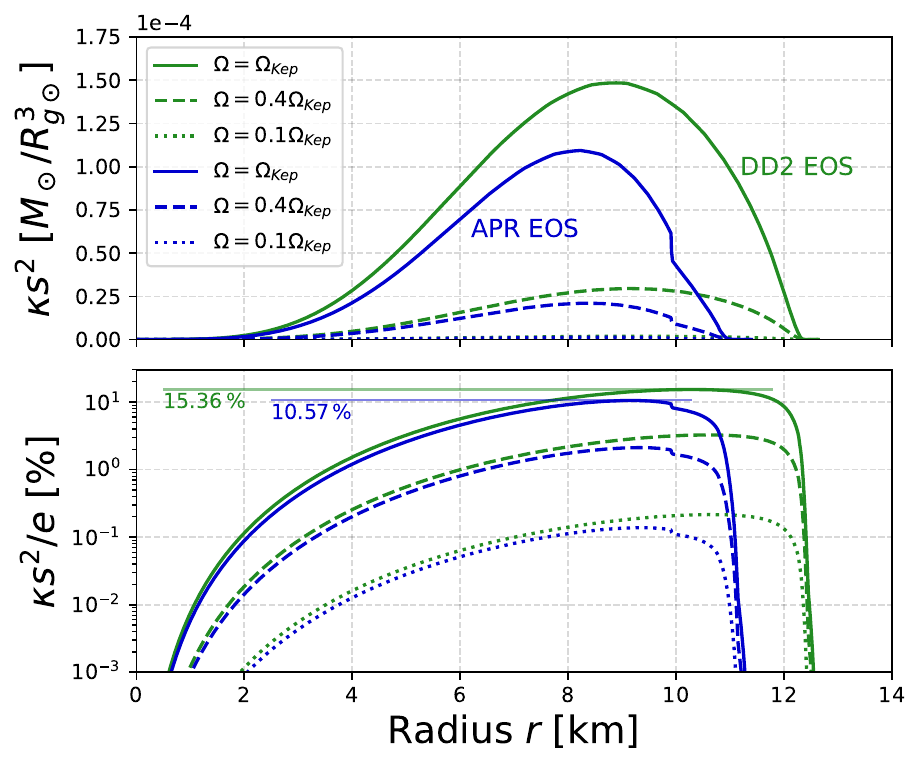}
    \caption{\textbf{Upper panel:} The magnitude of the spin-contribution $\kappa s^2$ to the effective energy density (see \eqref{eq:effective_quantities}), as given by \eqref{sec:uniformly-rotating-neutron-stars:eq:spin-density-extended-body-ns} (with $\theta = \frac{\pi}{2}$). The green and blue lines correspond to radial profiles of two neutron stars with a central density of $\rho_c=4\rho_{sat}$ and different equations of state, DD2 \cite{Hempel:2009mc} and APR \cite{Schneider:2019vdm}, respectively. Solid, dashed and dotted lines correspond to different rotational rates $\Omega$ for each star, given relative to the Keplerian rotation rate $\Omega_{Kep}^2=M/R^3$ of the corresponding star without torsion. \textbf{Lower panel:} Same as above, but this time we show the contribution of the spin effects $\kappa s^2$ relative to the total energy density $e$. The relative contributions can reach the order of $10\,\%$. The maximal relative contribution are marked explicitly for the stars at the Keplerian limit. Note that the values for $\kappa s^2$ were obtained non self-consistently using the values of the corresponding NS without torsion.
    }
    \label{fig:uniformly-rotating-neutron-stars:magnitude-of-spin-contributions-radial-plot}
\end{figure}

We now perform an order of magnitude estimation of the expected torsion effects in neutron stars. We use \eqref{sec:uniformly-rotating-neutron-stars:eq:spin-density-extended-body-ns} to estimate the magnitude of the effects as a function of the coordinate radius $r$. We assume different constant rotation rates $\Omega$, relative to the Keplerian velocity $\Omega_{Kep}^2 = M/R^3$ of the corresponding NS without torsion effects. We further consider the torsion effects on the equatorial plane ($\theta = \frac{\pi}{2}$), where they are expected to be largest. It must be noted that this estimate is not self-consistent. We simply computed the expected strength of the torsion effects according to \eqref{sec:uniformly-rotating-neutron-stars:eq:spin-density-extended-body-ns} for a non-rotating neutron star and without torsion effects. Using the known values of pressure and energy density, we then computed $s^2$ for every radial position inside of the neutron star. \\
~
In \autoref{fig:uniformly-rotating-neutron-stars:magnitude-of-spin-contributions-radial-plot}, we show the absolute and relative magnitude of the rotation-induced torsion effects $\kappa s^2$. The DD2 \cite{Hempel:2009mc} and APR \cite{Schneider:2019vdm} EOS are chosen as representatives of a stiff and a soft EOS, respectively. The torsion effects are zero near the centre of the star and the rise monotonically until they reach a maximum at roughly $\sim 70\%$ of the NS radius. The torsion effects then decrease in strength until they reach zero at the NS radius. This can be explained using \eqref{sec:uniformly-rotating-neutron-stars:eq:spin-density-extended-body-ns}. For constant rotation rate $\Omega$, the torsion effects rise roughly proportional to $r^4$. At the same time, the pressure $P$ and energy density $e$ of the NS matter decrease, until they reach zero at the NS surface. This leads to a ``sweet spot'' between radius of the star and falling energy density, where the rotation-induced torsion effects are strongest. The lower panel of \autoref{fig:uniformly-rotating-neutron-stars:magnitude-of-spin-contributions-radial-plot} shows that the relative contributions reach roughly $\sim 15\,\%$ at $\sim 90\,\%$ of the NS radius for the DD2 EOS, and $\sim 10\,\%$ at $\sim 80\,\%$ of the NS radius for the APR EOS. The precise number and radius will depend on the chosen EOS and radial matter distribution of the star. At these radii, one should expect to see the largest changes in the neutron star structure due to rotation-induced torsion effects. Also, the strength of the torsion effects increases when increasing the rotation rate, as expected. \\
Relative torsion effects on the order of $10-15\,\%$ are considerable and should significantly affect the internal composition of neutron stars. While effects might be small for slowly rotating NS, they might be highly relevant for fast rotating neutron stars. This could especially be the case in binary neutron star mergers, where the post-merger remnants are predicted to rotate at velocities close to the mass-shedding limit with roughly Keplerian rotation velocities, see \cite{Shibata:2019wef}. Thus, if torsion effects can be sourced through rotation, we expect that the impact on rotating neutron stars is non-negligible. We will discuss this point further in the next section.

\section{Results \label{sec:Results}} 
\subsection{Numerical Setup and Methods} \label{subsec:numerical results}

\noindent
We solve the TOV equations \eqref{eq:TOV_collection-a-eq}-\eqref{eq:TOV_collection-P-eq} numerically using the public code $\mathsf{FBS-Solver}$, developed by the authors of \cite{Diedrichs:2023trk}. It uses a Runge-Kutta-Fehlberg solver to integrate the equations. Our version of the code – including the data used to make all figures – can be found online here \cite{Jockel-Menger2024}. \\
We implemented the polytropic model for the spin density \eqref{eq:polytropic-spin-density-model} ($s^2=\beta P^\gamma$) in our code. One could also study the constant spin-density model \eqref{eq:constant-spin-density-model} ($s^2=s^2_0$) if one takes $\gamma=0$ and sets $\beta \equiv s_0^2$. \\
We do not solve the model for the spin fluid without spin polarization \eqref{eq:realistic-spin-density-model} ($s^2=\frac{1}{8}(\hbar n)^2$). This is because any contributions of the spin-density will not significantly affect solution of \eqref{eq:TOV_collection-a-eq}-\eqref{eq:TOV_collection-P-eq} at the densities present in a NS (see the discussion in \autoref{subsec:analytical-results}). Another way of motivating this, is by considering the units $c = G = M_\odot = 1$ used in this work. In these units, the Planck mass is $M_p = \sqrt{\hbar} \approx 1.1 \times 10^{-38}$. This directly implies that $\hbar \approx 1.2 \times 10^{-76}$. All terms in \eqref{eq:realistic-spin-density-model} related to spin contributions are proportional to $\hbar^2$. \LM{The values of the torsion terms will therefore be very small and difficult to resolve numerically. It can be concluded that likely no effects will be visible in any numerical solution in the context of neutron stars consisting of fermionic particles.} \\
Additionally, we consider the case where the torsion effects are not sourced by microphysical spin but rather by macroscopic angular momentum (see the discussion in \autoref{sec:uniformly-rotating-neutron-stars}). \\

To model the neutron matter, we use the DD2 equation of state (with electrons) \cite{Hempel:2009mc} and the APR EOS \cite{Schneider:2019vdm}. Both are taken from the CompOSE database \cite{Typel:2013rza}. We take the DD2 and APR EOS as representatives of a stiff and soft EOS, respectively. They were also chosen because they are widely used by a number of groups and thus are well known in the literature.

\subsection{Static Neutron Stars} \label{subsec:static-neutron-stars}

\begin{figure}
\centering
    \includegraphics[width=0.495\textwidth]{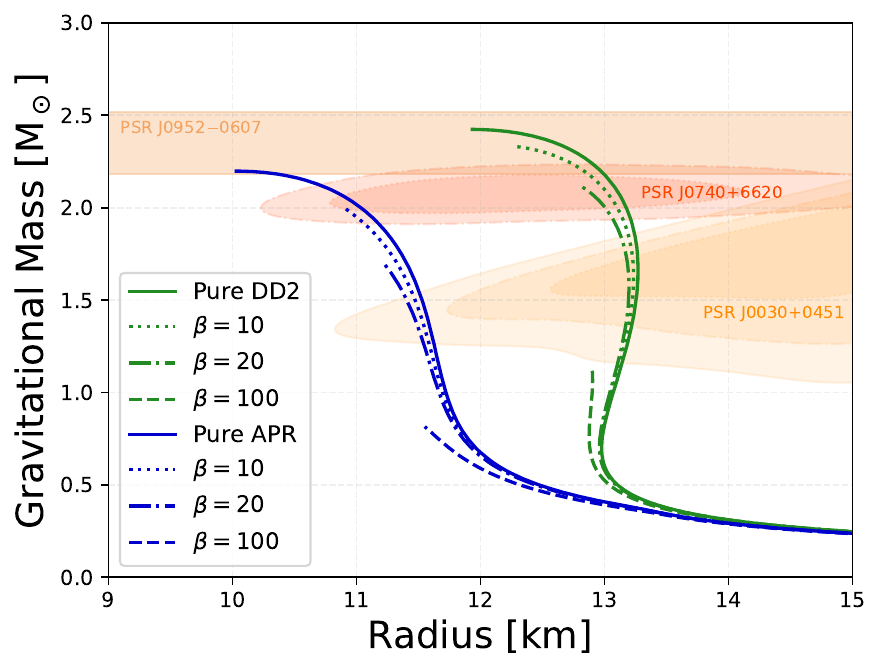}
    \caption{
    Mass-radius relations of static neutron stars for the DD2 \cite{Hempel:2009mc} (green lines) and APR \cite{Schneider:2019vdm} (blue lines) equation of state with varying torsion parameters $\beta$ and $\gamma=2$. The shaded regions indicate measurement constraints from different pulsars. The general trend is that an increase in torsion effects (larger $\beta$) tends to decrease both mass and radius. Deviations from the no-torsion configuration become larger with increasing $\beta$. There also exists a cutoff point after which there are no stable configurations with higher mass. The highest possible mass also decreases when increasing the torsion coupling strength.
    }
    \label{fig:static-neutron-stars:mr-relation}
\end{figure}

In this section, we investigate the effects that torsion sourced by micro-physical spin has on the properties of neutron stars. To that end, we use the polytropic model \eqref{eq:polytropic-spin-density-model} to gauge the nature and the order of magnitude of these effects. We chose different values for the parameters $\beta$ and $\gamma$, which encode information about the coupling strength between matter and torsion. For now, we consider values ($\beta$,$\gamma$) that are suitable to gauge the  general effects and to explore the parameter space. In the next section, we will make physically motivated choices for these parameters using the naive spin-density model that we derived for rotating stars. \\

In \autoref{fig:static-neutron-stars:mr-relation}, we show mass-radius curves of neutron stars for varying values of the $\beta$ parameter and $\gamma=2$. We use the DD2 and APR EOS to model the nuclear matter. The shaded regions mark observational constraints from measurements of the pulsars PSR J0952\ensuremath{-}0607 \cite{Romani:2022jhd}, PSR J0030\ensuremath{+}0451 \cite{Vinciguerra:2023qxq}, and PSR J0740\ensuremath{+}6620 \cite{Riley:2021pdl}. Since we use the polytropic model ($s^2=\beta P^\gamma$, \eqref{eq:polytropic-spin-density-model}), $\beta$ directly correlates to the strength of torsion effects. Increasing $\beta$ leads to NS configurations with lower mass and radius over all. This is independent of the equation of states used. There also exists a cutoff point at a specific maximum mass, after which there are no stable configurations. \\
The reason for this cutoff is the singularity that arises in \eqref{eq:polytropic-spin-density-model} when the denominator reaches zero, i.e. when
\begin{align}
    1 \stackrel{!}{=} \kappa \beta \gamma P^{\gamma -1}. \label{sec:results:eq:critical-density-condition}
\end{align}
With this equation, we can obtain a critical density using a given equation of state where $\rho (P)$ is known. For example for the polytropic EOS $P = K \rho^\Gamma$ ($K,\Gamma \in \mathbb{R}$), we obtain the critical density analytically:
\begin{align}
   \rho_{crit}= K^{-1/\Gamma} \left( \kappa \beta \gamma \right)^{1/\Gamma(1-\gamma)} \: . \label{sec:results:eq:critical-density-polytrope}
\end{align}
The maximal possible central density a neutron star can reach in the presence of torsion is thus limited. This also implies the existence of a highest possible neutron star mass -- where the NS has a central density slightly below the critical density $\rho_{crit}$. \\
For the cases without torsion, it is well known that the NS of maximal gravitational mass marks the last configuration that is stable to linear radial perturbations. This constraint should also apply to stars in Einstein--Cartan gravity. In addition, the constraint due to the critical density will also apply. This can be seen in \autoref{fig:static-neutron-stars:mr-relation} for the lines with $\beta > 0$. There, the point of highest mass corresponds exactly to the point where the critical density is reached. \\
This finding motivates us to further investigate the behaviour of the central density for different choices of $\beta$ and $\gamma$. We discuss this aspect further in the following paragraphs. \\

\begin{figure}
\centering
    \includegraphics[width=0.49\textwidth]{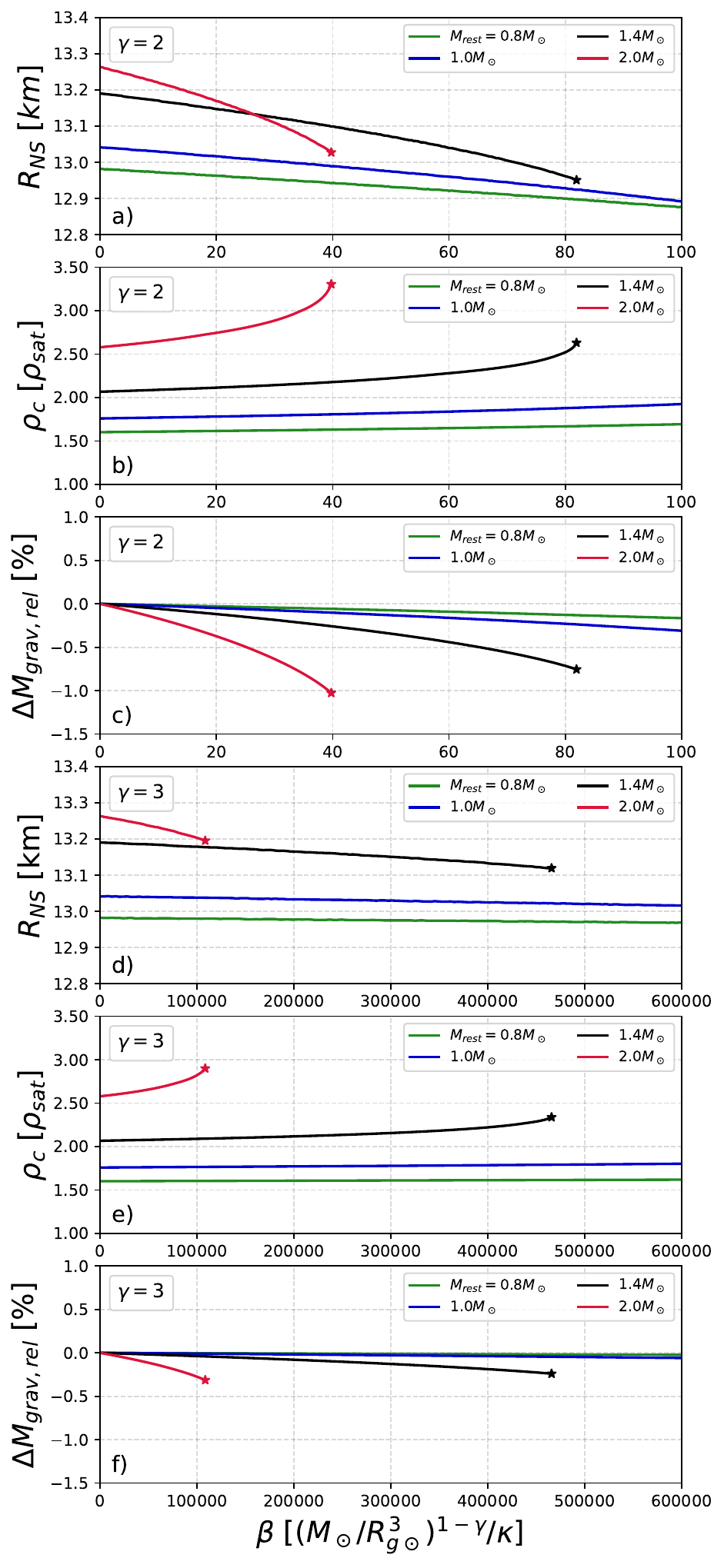}
    \caption{The radius (panels \textbf{a)} and \textbf{d)}), central density (panels \textbf{b)} and \textbf{e)}) and relative change in gravitational mass $M_\mathrm{grav}$ (panels \textbf{c)} and \textbf{f)}) for neutron stars of constant restmass $M_\mathrm{rest}$ as a function of $\beta$ and for different $\gamma = \{2,3\}$. Stars mark the points where the critical density $\rho_{crit}$ is reached and the NSs become unstable. With increasing $\beta$, the radius decreases, $M_\mathrm{grav}$ decreases and the central density increases. This also implies a higher binding energy since more matter is located deeper in the gravitational well. All effects are more pronounced for $\gamma=2$ compared to $\gamma=3$. All calculations were performed for the DD2 EOS \cite{Hempel:2009mc}.}
    \label{fig:static-neutron-stars:radius-rho-mass-beta-relation}
\end{figure}

In \autoref{fig:static-neutron-stars:radius-rho-mass-beta-relation} we show the radius $R_{NS}$, the central density $\rho_c$ and relative change in gravitational mass $M_\mathrm{grav}$ for neutron stars of constant restmass $M_\mathrm{rest}$ while varying $\beta$ and $\gamma = \{2,3\}$. All calculations were performed using the polytropic spin-density ansatz ($s^2=\beta P^\gamma$) and for the DD2 EOS \cite{Hempel:2009mc}. In all panels, the star $\star$ symbols mark the point at which the critical density is reached and no stable static solution exist (see the discussion of \autoref{fig:static-neutron-stars:mr-relation}). We investigate here how increasing torsion affects neutron stars of constant restmass. This can be seen as equivalent to taking a neutron star with given restmass without torsion effects, and then gradually increasing the torsion strength. The strength is quantified by the parameter $\beta$. \\

In panels \textbf{a)} and \textbf{d)}, we show the change in the NS radius when increasing torsion effects. Independent of $\gamma$, we see a decrease in the radius for increasing $\beta$. The radius change is on the order of $100-300\,m$. The smallest radius is reached at the point where the central density of the star reaches $\rho_{crit}$. \\
Panels \textbf{b)} and \textbf{e)} show the central density $\rho_c$ as a function of $\beta$. The central density rises for increasing $\beta$ until it reaches the critical density. The point where it reaches $\rho_{crit}$ is marked by a $\star$ symbol. This happens independently of $\gamma$, merely the specific value of $\rho_{crit}$ and $\beta$ change. The changes in density are of the order of $0.5-1$ times the nuclear saturation density. For $\gamma=3$, the critical central densities reach lower values and are located at larger values for $\beta$. When the equation of state in known, the concrete value of $\rho_{crit}$ can be computed using \eqref{sec:results:eq:critical-density-condition}. From this, we also see that the critical density is inversely proportional to $\beta$. The reason why the central density rises for larger $\beta$ in the first place is due to the repulsive nature of torsion. When considering the effective energy density $e_\mathrm{eff} = e - \kappa s^2$ (see \eqref{eq:effective_quantities}), we see that torsion has the effect of reducing it. This enables a given neutron star with a fixed restmass to support higher central densities, compared to an equal-restmass neutron star without torsion, while still being stable. A neutron star in EC gravity will thus have a higher central density as their counterpart with equal restmass in general relativity. This is can be seen especially well for the cases $M_\mathrm{rest}=1.4\,M_\odot$ and $M_\mathrm{rest}=2\,M_\odot$, where the case $\beta=0$ corresponds to the GR case. \\
Another way to read panels \textbf{b)} and \textbf{e)} is as follows. The star symbols mark the last stable NS configurations which have central densities just below $\rho_{crit}$. We can construct a curve between all these critical points. This curve will then mark a boundary between stable and unstable NSs. When increasing the NS mass by e.g. accretion, the given central density will also increase. For a given $\beta$, the NS will rise vertically in the diagram to higher $\rho_c$, until it reaches the boundary $\star$-curve. This also implies a maximum possible NS mass for every given $\beta$. \\
In panels \textbf{c)} and \textbf{f)}, we show the relative change in the gravitational mass. Because the restmass is held constant, this also corresponds to the change in binding energy $E_\mathrm{bind} := M_\mathrm{rest} - M_\mathrm{grav}$. One can see a clear decrease of $M_\mathrm{grav}$ with higher torsion strength on the order of up to $1\,\%$. This is consistent with the increase in central density and can be interpreted as follows: Stronger torsion effects allow for higher densities, which enables matter to be concentrated deeper in the gravitational well of the NS. This also directly translates to a higher gravitational binding energy. The higher binding energy could have observational consequences in the scenario of rapid collapse, where potential energy is released and converted into binding energy and radiation. \\

We return to the notion of the critical density and explore some of its consequences in the following. By definition, the critical central density $\rho_{crit}$ is associated with a singularity in $P'$ (see \eqref{sec:results:eq:critical-density-condition}). The point where the pressure derivative becomes singular thus also marks the point of maximal/critical density. For the sake of our argument, we assume that torsion effects are sufficiently strong to put $\rho_{crit}$ at densities achievable within NSs (e.g. via a large value of $\beta$). One could then ask what happens if a small amount of mass $\Delta M$ is added to a neutron star. \\
We first consider the case of a typical stable NS with a central density smaller then the critical density. Adding matter $\Delta M$ will lead to an increase in the central density. This is the case for all typical EOSs because the gravitational mass is a monotonous function of the central density and vice-versa (we here ignore exotic EOS behaviour like e.g. twin stars). Matter can be added incrementally until the central density reaches just below $\rho_{crit}$. \\
Now, what would happen if we add a further amount of mass $\Delta M$ to such a NS? The differential equation for the pressure \eqref{eq:polytropic-spin-density-model} predicts that surpassing the critical density leads to a sign change in the derivative of the pressure. The pressure would hence increase as a function of the radius. This is a clear sign that increasing the central density beyond $\rho_{crit}$ makes the NS unstable. After becoming unstable, the neutron star can either migrate to a configuration with lower central density via mass-shedding or internal re-arranging, or it can collapse into a black hole. The increase in pressure when above the critical density might also hint to the NS exploding instead. Precise numerical simulations are necessary to decide which scenario is realised. However constructing a numerical model which can deal with the arising non-equilibrium system goes beyond the scope of this paper. We leave this for future work. \\

In the above paragraph we discussed what might happen when the central density of a NS surpasses the critical density. Whether the critical density can be reached in realistic scenarios is another question entirely. As an example, we assume that torsion is sourced by microphysical spin and that the spin density is given by \eqref{sec:theory:eq:spin-fluid-eos-fermions}. The pressure inside of a NS then evolves according to \eqref{eq:realistic-spin-density-model}. The equation becomes singular when
\begin{align}
    1 - \frac{\kappa \hbar^2}{4} n \frac{\partial n}{\partial P} \stackrel{!}{=} 0 \: .
\end{align}
We can now estimate the necessary number density for this to happen. We do this for a fluid of fermions which obey the EOS of a Fermi gas:
\begin{align}
    P = \frac{(3\pi^2)^{2/3} \hbar^2}{5 m_\mathrm{f}} n^{5/3},
\end{align}
were $m_\mathrm{f}$ is the mass of the fermion (e.g. neutrons or quarks). This should suffice as an order of magnitude estimation. Under these assumptions, we find a critical number density (in SI units) for fermions:
\begin{align}
    n_{crit} =  \frac{64 \pi^4}{3} (m_\mathrm{f} c^2 \kappa)^{-3} \: .
\end{align} 
For neutrons one obtains the (ridiculous) number density $n_{crit} = 3.27 \times 10^{157} m^{-3}$. It lies far beyond the scope of any known phenomenon. This essentially eliminates any chance of investigating the critical density in the context of neutron stars for this microphysical spin density model (\eqref{sec:theory:eq:spin-fluid-eos-fermions}). Note that this result should be taken with great caution since at these length scales and densities, quantum mechanical effects and gravitational effects will without doubt both be relevant. \\

However microscopical spin is not the only way how torsion might affect compact objects. In the next section, we consider torsion effects induced from macroscopic rotation and whether these might leave noticeable imprints on the structure of neutron stars.

\subsection{Rotating Neutron Stars} \label{subsec:rotating-neutron-stars}

In this section, we investigate the effects that torsion sourced by macrophysical rotation has on the properties of neutron stars. The focus lies on getting an (upper) estimate for the expected magnitude and scale of rotation-induced torsion. We also discuss some physical consequences of our analytical model for rotation-induced torsion (derived in \autoref{sec:uniformly-rotating-neutron-stars}) in the case of neutron stars that spin up due to the accretion of angular momentum. \\

To get an upper estimate for the torsion effects due to rotation, we approximate the rotating model for the spin density \eqref{sec:uniformly-rotating-neutron-stars:eq:spin-density-extended-body-ns} using a similar form to the polytropic model (\ref{eq:polytropic-spin-density-model}) as
\begin{align}
    s^2 = \beta_{rot} (e+P)^2 \ , \qquad \beta_{rot} = R^4 \Omega^2 \Gamma^4 \ . \label{sec:results:eq:rotating-ns-simplified-rotating-model}
\end{align}
Here $R$ is the neutron star radius, $\Omega$ is the angular frequency and $\Gamma = 1/\sqrt{1-R^2\Omega^2}$ is the Lorentz factor. $\beta_{rot}$ is an upper bound for the torsion effects, because $\beta = r^4 \Omega^2 \Gamma^4$ (see \eqref{sec:uniformly-rotating-neutron-stars:eq:spin-density-extended-body-ns} \LMtwo{with $\theta = \frac{\pi}{2}$}) is maximal at $R$. \LMtwo{Interestingly, the weaker torsion effects near the rotation axis (where $\cos(\theta)^2 \ll 1$) might contribute to a more prolate shape of the rotating NS. This is because torsion effects tend to reduce the NS radius and this effect will be strongest at the equator.} \\
Using the model in \eqref{sec:results:eq:rotating-ns-simplified-rotating-model}, the equation of motion for the pressure becomes
\begin{align}
        P' &= - \frac{(e+P - 2 \kappa \beta_{rot} (e+P)^2)}{(1 - 2 \kappa \beta_{rot} (e+P)(1+\partial e/\partial P) )} \, \frac{\alpha'}{\alpha\,} \: ,
        \label{sec:results:eq:rotating-ns-simplified-rotating-model-pressure-ode}
\end{align}
where $\partial e/\partial P$ is related to the local speed of sound $c_s$ through $\partial e/\partial P = 1/c_s^2$. In our code, we solve the TOV equations \eqref{eq:TOV_collection} and use \eqref{sec:results:eq:rotating-ns-simplified-rotating-model-pressure-ode} for the pressure. The NS radius $R$ in \eqref{sec:results:eq:rotating-ns-simplified-rotating-model} is obtained iteratively: As an initial guess, we take $R$ as the radius of the non-rotating NS with a given central density. We then compute $\beta_{rot}$ and from this we get a new configuration with a new radius $R$, which will be smaller than the previous one. We then use the new $R$ to update the value of $\beta_{rot}$. $\beta_{rot}$ will be slightly smaller and subsequently lead to a slightly larger $R$. This is repeated until the values converge. \\ 
It is also possible to derive $P'$ using the full spin density in \eqref{sec:uniformly-rotating-neutron-stars:eq:spin-density-extended-body-ns}. More details are given in Appendix \ref{sec:appendix:full-pressure-ode-for-rotational-spin-density}. We do not use this here because we are mainly interested upper bounds and the order of magnitude of torsion effects due to rotation. \\

\begin{figure}
\centering
    \includegraphics[width=0.495\textwidth]{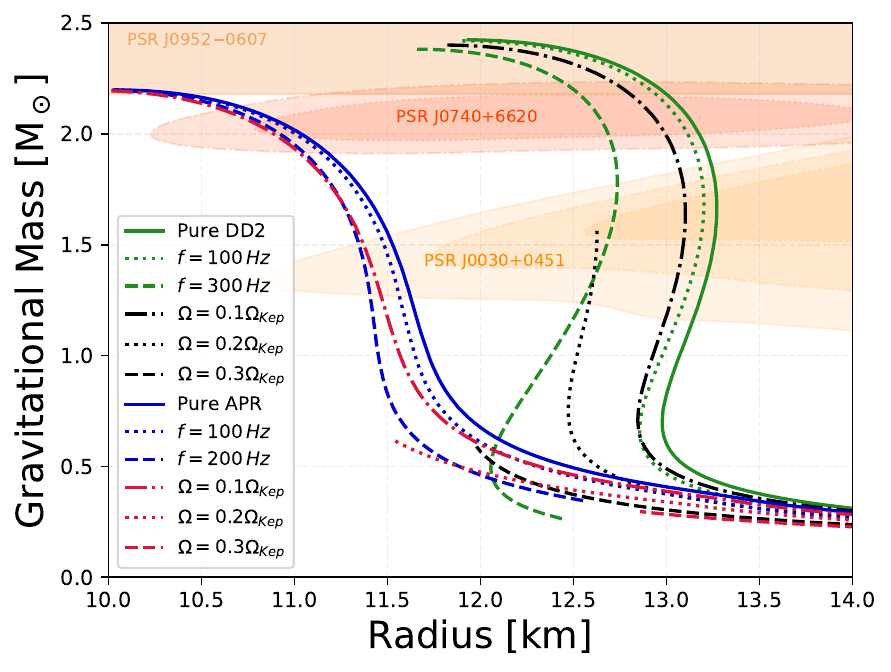}
    \caption{
        Mass-radius relations of different NSs with rotation-induced torsion effects computed using the ``upper-limit'' prescription from \eqref{sec:results:eq:rotating-ns-simplified-rotating-model} and \eqref{sec:results:eq:rotating-ns-simplified-rotating-model-pressure-ode}. Green and black lines correspond to NS with the DD2 EOS\cite{Hempel:2009mc}, blue and red to the APR \cite{Schneider:2019vdm}) EOS. All lines correspond to either NSs with constant rotation frequencies $\mathrm{f}=\Omega/2\pi$ or with fractions of the Keplerian rotation rate $\Omega_{Kep}=\sqrt{M/R^3}$. The shaded regions indicate measurement constraints from different pulsars. The general trend is that an increase in torsion effects (larger rotation rates) tends to decrease both mass and radius. Deviations from the no-torsion configuration become larger with increasing rotation. Some lines show cut-off points where the central density surpasses the critical density. This is an artifact of the simplified prescription (\ref{sec:results:eq:rotating-ns-simplified-rotating-model}).
    }
    
    \label{fig:rotating-neutron-stars:mr-relation-model-e+P}
\end{figure}

In \autoref{fig:rotating-neutron-stars:mr-relation-model-e+P}, we show mass-radius curves of NSs with rotation-induced torsion effects computed using the ``upper-limit'' prescription from \eqref{sec:results:eq:rotating-ns-simplified-rotating-model} and \eqref{sec:results:eq:rotating-ns-simplified-rotating-model-pressure-ode}. We use the DD2 and APR EOS to model the nuclear matter. All lines correspond to either NSs with constant rotation frequencies $\mathrm{f}=\Omega/2\pi$ or with fractions of the Keplerian rotation rate $\Omega_{Kep}=\sqrt{M/R^3}$. The shaded regions mark observational constraints from measurements of the pulsars PSR J0952\ensuremath{-}0607 \cite{Romani:2022jhd}, PSR J0030\ensuremath{+}0451 \cite{Vinciguerra:2023qxq}, and PSR J0740\ensuremath{+}6620 \cite{Riley:2021pdl}. \\
As expected, higher rotation rates lead to stronger torsion effects (quantified by $\beta_{rot}$) and leave a larger impact on the NS mass and radius than slower rotation rates. As found previously for microphysical spin, rotation-induced torsion effects also lead to a decreased radius and gravitational mass. We find that even for modest rotation rates, the change in radius can be up to several hundred meters. However it must be noted that we do not compute rotation of the NSs self-consistently. In reality, the centrifugal forces due to rotation would act to increase the NS radius. We discuss this aspect in detail later. \\
With our present model, we are able to probe rotation frequencies of up to a few hundred Hz or up to $20-30\,\%$ of the Keplerian rotation rate. This limitation comes from our simplified approach to model rotational torsion effects using \eqref{sec:results:eq:rotating-ns-simplified-rotating-model} and \eqref{sec:results:eq:rotating-ns-simplified-rotating-model-pressure-ode} (also see the discussion in \autoref{sec:uniformly-rotating-neutron-stars}). \autoref{fig:rotating-neutron-stars:mr-relation-model-e+P} also shows cut-off points, where the central density would surpass the critical density (see e.g. the green dashed line for $\mathrm{f}=300\,$Hz). This is an artifact of the simplified prescription (\ref{sec:results:eq:rotating-ns-simplified-rotating-model}) - (\ref{sec:results:eq:rotating-ns-simplified-rotating-model-pressure-ode}): To get an upper bound, we introduced an effective torsion coupling strength parameter $\beta_{rot}$, which is constant throughout the NS. But we know from \eqref{sec:uniformly-rotating-neutron-stars:eq:spin-density-extended-body-ns} that rotation-induced torsion effects should be zero along the axis of rotation and thus a critical density should not exist. The results from \autoref{fig:rotating-neutron-stars:mr-relation-model-e+P} should therefore be understood with regards to its general trend with conservative error bars in mind. \\

If they exist, rotation-induced torsion effects will play a role as soon as some amount of angular momentum $\Delta L$ is added to a non-rotating neutron star. This could happen e.g. through non-spherical accretion of matter onto the NS. The NS would then start to rotate and the torsion effects would decrease the radius of the NS by some amount. Due to the conservation of angular momentum, $L \propto R^2 \Omega M$, the decrease in radius also leads to a spin-up of the neutron star. Hence, the NS quantities before the spin-up (subscript $1$) and after (subscript $2$) change as
\begin{align}
    R_2 < R_1 \:\: , \:\: M_2 < M_1 \:\: , \:\: \Omega_2 = \Omega_1 \left( \frac{R_1}{R_2} \right)^2 \frac{M_1}{M_2}.
\end{align}
Note that here, $M_{i}$ denotes the gravitational mass. The restmass of the NS stays constant during spin-up or spin-down. Following from this, we also find that the Lorentz factors obey $\Gamma_2 > \Gamma_1$. In addition, the central densities and pressures increase, such that $e_1+P_1 < e_2 + P_2$ (see \autoref{fig:static-neutron-stars:radius-rho-mass-beta-relation}). \\
\LM{For the rotation-induced model of the spin-density scalar $s^2$ considered in this work (\eqref{sec:results:eq:rotating-ns-simplified-rotating-model-pressure-ode}), we find that $s^2_2 > s^2_1$.} This means that the strength of the torsion effects will increase further. Assuming that the radius is only affected by torsion effects, this leads to an uncontrolled and self-reinforcing spin up (a ''spin-up death spiral''), where the radius continuously decreases and the rotation rate increases. Eventually this leads either to the collapse of the NS into a black hole, to tidal disruption due to quick rotation, or to instability due to passing the critical central density, depending which limit is reached first. \\

However, there is one effect which could counteract this trend: The centrifugal force due to rotation. While stronger torsion effects reduce the NS radius, stronger centrifugal forces increase it. As a result, there are a number of possible scenarios when considering centrifugal forces and torsion effects together. They depend on which effect is larger and whether an equilibrium between both effects can exist. \\
If the torsion effects always dominate over the centrifugal forces, then the spin-up death spiral ensues. If centrifugal forces always dominate, then the NS would settle into an equilibrium configuration, dominated by the outward centrifugal force, albeit with smaller radius compared to the zero-torsion case. \\

Another option is that both effects reach equal magnitudes at some point. If torsion effects dominate for small rotation rates, the spin-up would continue until the centrifugal forces catch up in strength to stop further spin-up. This implies that any neutron star will reach a characteristic minimal rotation rate. The rotation rate would depend on the NS mass and equation of state. \\
If torsion effects dominate for larger rotation rates, this would lead to a characteristic maximal stable rotation rate. Beyond this rate, all NSs would be caught in a spin-up death spiral. \\

It is difficult to say, based on our results alone, which of the two effects will dominate and which scenarios, if any, are realised in nature. Note also that the characteristic spin-up/spin-down timescale due to torsion is not known. If this timescale is significantly shorter than the typical NS lifetime, this could set strong constraints on torsion effects due to rotation, or even exclude them. Further research on fully self-consistent solutions for rotating neutron stars in EC gravity is needed to settle this debate.
\section{Conclusions \label{sec:conclusions}}
In this work, we investigated how spacetime torsion affects the structure of neutron stars (NSs). We modeled torsion as a non-propagating property of spacetime in the framework of Einstein--Cartan gravity. In this theory, torsion leads to an additional coupling between spacetime and angular momentum. We derived the corresponding Tolman-Oppenheimer-Volkoff (TOV) equations, which describe the structure of static and spherically symmetric neutron stars. We then used the TOV equations to model the coupling of different sources for torsion to gravity. \\
We considered two physically motivated choices as sources for torsion effects: Microphysical spin and macroscopic angular momentum. For microphysical spin, we considered a model for the spin-density of a fluid of fermions (spin-fluid) and a phenomenological ``polytropic'' model. For macroscopic angular momentum, we derived a model where torsion is sourced by the macroscopic rotation of the neutron star. We then investigated these different models analytically and numerically. \\

For the fermion-fluid, we found that the expected effects are very small and likely negligible inside of neutron stars. We therefore investigated the phenomenological model for the microscopic spin density to gauge the general effects of microscopic torsion. For this, we considered neutron stars with constant restmass. We varied the effective coupling strength of microphysical torsion to see how torsion affects neutron stars in general. The coupling strength has no inherent physical meaning apart from revealing the expected phenomenology of torsion. \\
We found that when including torsion effects, the gravitational mass and the radius of the NSs decrease, while the density increases. This can be understood as torsion leading to effective repulsive forces, corresponding to a negative contribution to the effective total energy density of matter-energy and torsion-energy (see Eq. (\ref{eq:effective_quantities})). As a result, more matter can be concentrated within the same volume because the additional torsion contributions can stabilize the matter at higher density compared to the case without torsion. \\
The torsion effects also lead to a critical bound for the central density, above which no stable NS can exist. Beyond this density, a NS will either adapt its structure, e.g. by shedding mass or internal re-arranging, or implode/explode. The presence of a critical density could therefore have strong implications on the merger and collapse of neutron stars. It should be noted however that when torsion is sourced from a spin-fluid, the expected order of magnitude of the critical density is likely far too high to be realised in any sensible physical scenario. \\

We also investigated in detail the case where torsion is sourced by macroscopic rotation. In section \ref{sec:uniformly-rotating-neutron-stars} we derived a number of estimates for the expected magnitude of torsion effects due to rotation. We derived explicitly the spin-density $s^2$, which quantifies torsion effects. $s^2$ was found to depend on the radius within the object, the rotation rate, and the energy density and pressure. We found that rotational torsion effects should be negligible for non-relativistic astronomical objects such as main sequence stars and planets. For rapidly rotating neutron stars however, the effects could be large and have an up to $15\,\%$ contribution to the overall energy density. This means that those effects cannot be neglected. In our numerical investigation, we found that rotation-induced torsion effects are zero along the rotational axis of the object and increase in strength with increasing axial distance. They fall off in strength when the matter densities become small near the NS surface.\\

Regarding the global quantities of NSs, rotational torsion effects are found to decrease the gravitational mass and the radius. The radius can be decreased by up to $900\,m$. This poses an interesting question regarding the balance of forces inside of rotating NSs. We performed a Gedankenexperiment where we added a small amount of angular momentum to an initially non-rotating NS and investigated the different possible outcomes. The torsion effects will decrease the NS radius but centrifugal forces will act to increase the radius. Depending on which effect dominates, this could lead to a runaway torsion-induced spin-up of neutron stars (we call it the ``spin-up death spiral''). \\
However, it should be noted that our analysis includes simplifying assumptions which enable us to give (rough) upper bounds on the strength of rotation-induced torsion effects. We therefore believe that it is necessary and interesting to consider a fully self-consistent model for rotation-induced effects. This is ongoing work by the authors. \\

Our results also are also relevant in the wider context of metric-affine theories of gravity. In these theories, the metric and the connection are independent degrees of freedom. Depending on the choice of the connection, non-propagating torsion and non-metricity can then appear and couple non-trivially to matter. Our work opens the way to study and constrain a wide array of metric-affine modifications of gravity using neutron stars. This can have a large impact on studying the physical viability of these classes of theories.

\begin{acknowledgments}
We thank the organizers of the Bad Honnef Physics School on Black Holes (2022) for providing the environment where many of the ideas for this project were kickstarted. We thank Adrià Delhom for providing feedback and proof reading of the manuscript. C.J. thanks Matteo Stockinger and the CRA group at the MPI for Gravitational Physics (Albert Einstein Institute) for discussions which helped to sanity-check some concepts in this work. \LM{We also thank the anonymous referee for comments which improved the quality of this work.}
\end{acknowledgments}
\appendix
\section*{Appendix \label{sec:appendix}}
\section{Components of angular-momentum density tensor} \label{sec:appendix:angular-momentum-tensor-of-extended-body}

Using the ansatz for the four velocity and metric described around Equations (\ref{sec:uniformly-rotating-neutron-stars:eq:extended-body-energy-momentum-tensor})--(\ref{sec:uniformly-rotating-neutron-stars:eq:extended-body-four-velocity}), the nonzero components of the energy-momentum tensor are:

\begin{subequations}
\label{sec:appendix:eq:angular-momentum-tensor-of-extended-body}
    \begin{align}
    T^{tt} &= (e+P) \Gamma^2 - P \: , \\
    T^{tx} &= -(e+P) \Gamma^2 r \Omega \sin(\phi) \: , \\
    T^{ty} &= (e+P) \Gamma^2 r \Omega \cos(\phi) \: , \\
    T^{xx} &= (e+P) \Gamma^2 r^2 \Omega^2 \sin^2(\phi) + P \: , \\
    T^{xy} &= (e+P) \Gamma^2 r^2 \Omega^2 \sin(\phi) \cos(\phi) \: , \\
    T^{yy} &= (e+P) \Gamma^2 r^2 \Omega^2 \cos^2(\phi) + P \: , \\
    T^{zz} &= P \: .
\end{align}
\end{subequations}
To compute the spin-density tensor $s^{\mu\nu}$ we first need to compute the components of the $\mathcal{M}^{\alpha \beta \gamma}$ tensor. We use \eqref{sec:uniformly-rotating-neutron-stars:eq:extended-body-four-position}--(\ref{sec:uniformly-rotating-neutron-stars:eq:extended-body-four-position-rotation-axis}) and the energy-momentum tensor form above. The components that contribute to the final spin-density tensor are the $\gamma = t$ components and are given by:

\begin{subequations}
    \begin{align}
    \mathcal{M}^{\alpha \alpha t} &= 0 \:\:\:\: \forall\, \alpha \in \{t,x,y,z\}\: , \\
    \mathcal{M}^{tx t} &= r \cos(\phi) \sin(\theta) (P - (e+P)\Gamma^2 ) \: , \\
    \mathcal{M}^{ty t} &= r \sin(\phi) \sin(\theta) (P - (e+P)\Gamma^2 ) \: , \\
    \mathcal{M}^{tz t} &= r \cos(\theta) (P - (e+P)\Gamma^2 ) \: , \\
    \mathcal{M}^{xy t} &= r^2 \Omega \Gamma^2 (e+P) \sin(\theta) \: , \\
    \mathcal{M}^{xz t} &= r^2 \Omega \Gamma^2 (e+P) \sin(\phi) \cos(\theta) \: , \\
    \mathcal{M}^{yz t} &= - r^2 \Omega \Gamma^2 (e+P) \cos(\phi) \cos(\theta) \: .
    \end{align}
    \label{eq:appendix:curly-m-tensor-components}
\end{subequations}

\section{Details on Gauging Away Components of the Spin-Density Tensor} \label{sec:appendix:details-on-gauging-away-components-of-the-spin-density-tensor}

The Dixon condition condition $L_{\mu\nu} P^\nu = 0$ and its pysical meaning was already explained in the main text, see \eqref{sec:uniformly-rotating-neutron-stars:eq:dixon-condition}. We here show how to use the Dixon condition to gauge away the $s^{t\mu}$-components of the spin-density tensor $s^{\mu\nu}=\mathcal{M}^{\mu\nu \alpha} n_\alpha$. To do this, we use the four-momentum of the extended body $P^\mu$ (see \eqref{sec:uniformly-rotating-neutron-stars:eq:extended-body-four-momentum-extended-body}) and the angular momentum tensor $L_{\mu\nu}$ (see \eqref{sec:uniformly-rotating-neutron-stars:eq:angular-momentum-tensor}). \\
For the given energy-momentum tensor (see \eqref{sec:appendix:eq:angular-momentum-tensor-of-extended-body}), the only nonzero component of $P^\mu$ is the $P^t$ (the others are zero due to the angle integrals over $\phi$). We can then impose the Dixon-condition $L_{\mu\nu} P^\nu = 0$. Because $P^t \neq 0$ we then get the following constraint equations:
\begin{align}
    L_{x\nu} P^\nu = L_{xt} P^t \stackrel{!}{=} 0 &\Longrightarrow L_{xt} = 0 \\
    L_{y\nu} P^\nu = L_{yt} P^t \stackrel{!}{=} 0 &\Longrightarrow L_{yt} = 0 \\
    L_{z\nu} P^\nu = L_{zt} P^t \stackrel{!}{=} 0 &\Longrightarrow L_{zt} = 0
\end{align}
That means that we can gauge away the $L_{t\mu}$-components using the Dixon condition and set them to zero. \\
We apply this fact to constrain the components of $s^{\mu\nu}=\mathcal{M}^{\mu\nu \alpha} n_\alpha$. The angular momentum tensor $L^{\alpha\beta}$ is related to the spin density tensor $s^{\mu\nu}$ via

\begin{align}
    L^{\alpha\beta} = \oint_{\partial U} \mathcal{M}^{\alpha\beta\gamma} d \Sigma_\gamma = \int \mathcal{M}^{\alpha\beta \gamma} n_\gamma dV_3 = \int s^{\mu\nu} dV_3 \: .
\end{align}
On the integration interval, The components of $s^{\mu\nu}$ have no sign change (i.e. are positive definite) and are nonzero. The integral bounds are also not symmetric. The integral of a certain component being zero then implies that the integrand must also be zero. When one component of $L^{\mu\nu}$ is zero this then implies that the corresponding component of $s^{\mu\nu}$ is zero as well. In our case this means that $s^{tx} = s^{ty} =s^{tz} = 0$. The only remaining components then are: $s^{xy}$, $s^{xz}$, $s^{yz}$ (also note that $s^{\mu\nu}$ is antisymmetric). For the ansatz chosen in the main text, some additional components of $s^{\mu\nu}=\mathcal{M}^{\mu\nu \alpha} n_\alpha$ are also zero. In the end, only the $s^{xy}$ contributes to the spin-density scalar.

\section{Full Pressure ODE for Rotational Spin Density} \label{sec:appendix:full-pressure-ode-for-rotational-spin-density}

We briefly show the derivation of the corresponding ODE for the pressure when using the spin density model in \eqref{sec:uniformly-rotating-neutron-stars:eq:spin-density-extended-body-ns} for the rotating matter configuration. We only consider the case on the equatorial plane ($\theta = \frac{\pi}{2}$) since there the effects are largest. To derive the ODE for the pressure, we insert \eqref{sec:uniformly-rotating-neutron-stars:eq:spin-density-extended-body-ns} into \eqref{eq:TOV_collection-P-eq} and take the radial derivative of $s^2$. We then use the chain rule as $de/dr=\partial e/\partial P \times dP/dr$, which is valid if $e=e(P)$. We rewrite $\partial e/\partial P = 1/c_s^2$ with the speed of sound $c_s$ and solve for $P'$. We finally obtain:

\begin{align}
    P' = \frac{\left[ 4\kappa s^2 \left( \frac{1}{r} + r \Omega^2 \Gamma^2 \right)  - \left( e + P - 2\kappa s^2 \right) \frac{\alpha'}{\alpha} \right]}{ \left( 1 - 2\kappa s^2 \frac{(1 + 1/c_s^2)}{(e+P)} \right)}
\end{align}

\noindent
This equation correctly captures the behaviour at $r \rightarrow 0$ where the torsion effects become zero ($s^2 \rightarrow 0$). Thus, a critical central density (see \autoref{subsec:static-neutron-stars}) does not exist for this model. However, this model has problems to produce consistent results at large $r$, because the first term in the numerator will grow unbounded and will force the pressure derivative to be positive. This leads to unphysical increasing pressures for large radii. This likely is a consequence of neglecting metric components and assuming a flat spacetime in the derivation of \eqref{sec:uniformly-rotating-neutron-stars:eq:spin-density-extended-body-ns} (see \autoref{sec:uniformly-rotating-neutron-stars} for the derivation).


\bibliographystyle{apsrev4-1}
\bibliography{biblio.bib}{}

\end{document}